\documentclass[final,10pt]{elsarticle}

\usepackage{amssymb}
\usepackage{amsmath}
\usepackage[margin=0.8in]{geometry}
\usepackage{lineno}
\usepackage{subfigure}
\usepackage{placeins}
\journal{Journal of Computational Physics}
\begin{document}

\begin{frontmatter}

\title{Adding slow magnetoacoustic mode to the HLL-type multi-state approximate Riemann solution}

 \author[label2,label3]{Fan Zhang}
    \author[label1,label4]{Stefaan Poedts}
  \author[label1]{Andrea Lani}

\affiliation[label2]{organization={Institute of Theoretical Astrophysics, University of Oslo},
             addressline={PO Box 1029 Blindern},
             city={Oslo},
             postcode={0315},
             country={Norway}}             
 \affiliation[label3]{organization={Rosseland Centre for Solar Physics, University of Oslo},
             addressline={PO Box 1029 Blindern},
             city={Oslo},
             postcode={0315},
             country={Norway}}

  \affiliation[label1]{organization={Centre for mathematical Plasma-Astrophysics, Department of Mathematics, KU Leuven},
             addressline={Celestijnenlaan 200B},
             city={Leuven},
             postcode={3001},
             country={Belgium}}

\affiliation[label4]{organization={Institute of Physics, University of Maria Curie-Sk{\l}odowska},
             addressline={Pl. M. Curie-Sk{\l}odowskiej 5},
             city={Lublin},
             postcode={20-031},
             country={Poland}}

\begin{abstract}
Multi-state HLL-type approximate Riemann solutions of ideal magnetohydrodynamics (MHD) typically assume that the medium within the Riemann fan is incompressible, and thus the slow magnetoacoustic mode cannot be included in their space-time  
wave configurations for the intermediate states. 
We propose a new strategy to design  multi-state HLL-type solutions, allowing for the medium within the Riemann fan to be compressible. In particular, for the complete seven-wave configuration of the MHD Riemann problem, we estimate the slow magnetoacoustic speeds and a longitudinal flow speed between the slow modes, whose compressibility is estimated based on the relations between fast, Alfv\'en, and slow speeds.
Then, following the conservation laws and Rankine-Hugoniot jump relations across the fast, Alfv\'en, and slow waves, we calculate all the intermediate states within the Riemann fan. Moreover, we discuss the solutions when certain wave modes degenerate,  ensuring well-posedness and smooth transitions between complete and degenerate wave configurations. Numerical simulations using a Finite Volume solver show that the proposed approximate Riemann solution is less diffusive than the classic HLLD scheme, particularly for slow mode waves. For example, in a 1D test case with a strong longitudinal magnetic field, the new scheme needs one order of magnitude fewer grid cells compared to the classic HLLD scheme to resolve all wave modes. 
\end{abstract}

\setcounter{figure}{-1}

\begin{graphicalabstract}
\end{graphicalabstract}

\setcounter{figure}{0}

\begin{highlights}
\item An "equivalent" propagation speed of compressible waves proposed for solving the intermediate states  
\item Flux conservation calculated separately for left and right nonlinear characteristics
\item Slow magnetoacoustic mode included in a multi-state HLL-type scheme 
\item Higher-resolution shown particularly when capturing slow magnetoacoustic waves
\end{highlights}

\begin{keyword}
approximate Riemann solution \sep magnetohydrodynamics \sep HLL-type scheme  \sep slow magnetoacoustic mode \sep Finite Volume method


\end{keyword}

\end{frontmatter}


\section{Introduction} \label{sec:intro}

 Hyperbolic conservation laws, such as the ideal magnetohydrodynamics (MHD) equations, are frequently solved or examined under the one-dimensional (1D) assumption using Finite Volume (FV) or Finite Difference (FD) frameworks. Consequently, the original multi-dimensional conservation laws are typically reduced to 1D partial differential equations (PDEs) before constructing numerical solutions. For example, we may write the 1D MHD equations as a first-order system
\begin{eqnarray}  \label{eq:1DMHD}
 \frac{\partial \mathbf{U}}{\partial t}+  \frac{\partial\mathbf{F}(\mathbf{U})}{\partial x}=\mathbf{0}.
 \end{eqnarray}
\noindent The conservative variables and flux functions are  
 \begin{eqnarray}\label{eq:govern} 
\mathbf{U}  =  \left(
\begin{array}{cc}
\rho  \\
\rho \mathbf{V}\\ 
E \\
\mathbf{B}
\end{array}     \right), \quad \text{and}
\quad \mathbf{F}(\mathbf{U}) =
 \left(
\begin{array}{cc}
\rho u\\
\rho u\mathbf{V}+P\mathbf{n}-B_x\mathbf{B}\\ 
u(E+P)-B_x(\mathbf{B}\cdot\mathbf{V}) \\
u\mathbf{B} - B_x\mathbf{V}  
\end{array}     \right),
\end{eqnarray}
\noindent where $\mathbf{n}=(1,0,0)^{\text{T}}$ in a 1D space. More specifically, $\rho$ is density, $\mathbf{V}=(u,v,w)^{\text{T}}$ is velocity, $\mathbf{B}=(B_x,B_y,B_z)^{\text{T}}$ is magnetic field, the total energy $E=\rho e + \frac{1}{2}\rho\mathbf{V}^2+\frac{1}{2}\mathbf{B}^2$ consists of internal energy, kinetic energy, and magnetic energy, and the total pressure $P=p+\frac{1}{2}\mathbf{B}^2$ includes plasma gas pressure and magnetic pressure. Here we use the ideal gas law $p=\rho e(\gamma -1)$ to close the equations, where $\gamma$ is the adiabatic index.
In addition, the 1D divergence constraint of the magnetic field is  
  \begin{eqnarray} \label{eq:Bn}
 \frac{\partial B_x}{\partial x}=0,
   \end{eqnarray}
\noindent which is naturally guaranteed in a 1D scenario, but is not equivalent to the multi-dimensional divergence constraint. We do not discuss the details related to the divergence constraint in this work, as they can be found in various references \cite{Evans_1988,BALSARA1999,Toth_2000,Dedner_2002,Balsara_2004}.

 Without going into a detailed explanation, we emphasize that an approximate Riemann solver can be used to provide numerical fluxes for the semi-discrete equations of Eq.~\eqref{eq:1DMHD} at a given time-step, i.e.,
\begin{eqnarray}  \label{eq:semi-discrete}
 \frac{\text{d} \mathbf{U}_i}{\text{d} t}=-  \frac{{\mathbf{F}}_{i+1/2}-{\mathbf{F}}_{i-1/2}}{\Delta x},
 \end{eqnarray}
 \noindent where $i$ denotes the grid point/cell on finite difference/volume discretizations, and subscripts $\pm 1/2$ denote that the numerical fluxes are calculated  at  the half-point/interface between the grid point/cell $i$ and its neighbouring grid points/cells $i\pm 1$.  The spatial distributions of the variables are typically calculated using piecewise polynomial approximations, but in this work we do not discuss this topic.

\begin{figure}[htbp]
 \centering
 \includegraphics[width=0.6\textwidth]{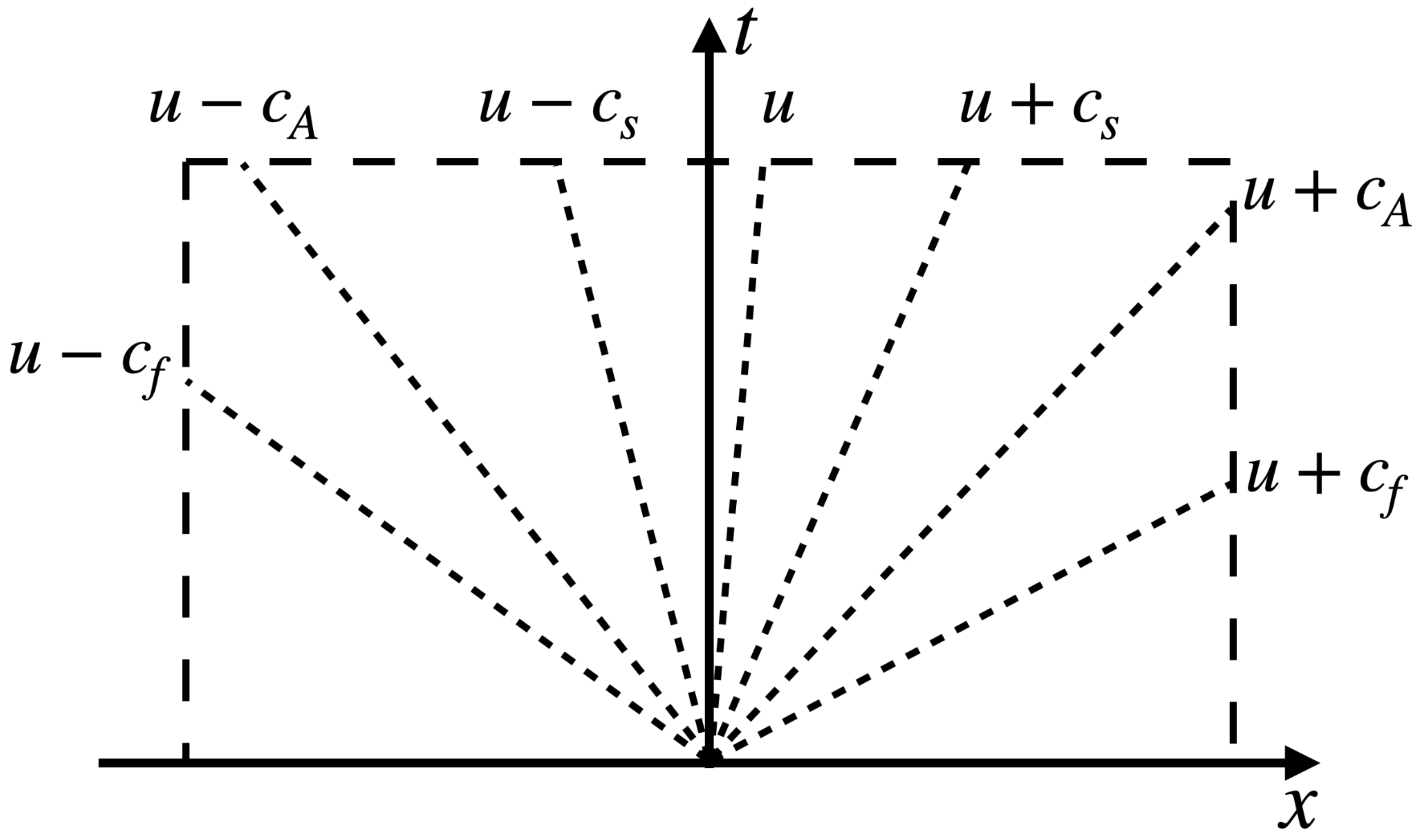}
 \caption{A 1D space-time schematic of eigenwave propagation in the MHD Riemann problem. Here, $u$ is the local flow speed, and $c$ denotes the eigenspeeds. The subscripts $f$, $A$, and $s$ denote fast, Alfv\'en, and slow modes and the corresponding 
 speeds.}
 \label{fig:MHD_modes}
\end{figure}

Originally developed to solve the gas dynamics equations, HLL-type (where HLL stands for Harten, Lax, van Leer) Riemann solvers \cite{Harten_1983,Toro1994} have also been extended to solve the MHD equations \cite{Gurski2004,Li2005,Miyoshi2005}, including relativistic MHD \cite{Mignone_2006,Mignone_2009,Mattia2021}. While a 1D MHD system has seven eigenwaves, as shown in Fig.\ref{fig:MHD_modes},
the general idea of HLL-type schemes is to assume a space-time  
wave configuration for the approximate solution of a Riemann problem. The assumed wave configuration consists of two or more eigenwaves of different speeds separating the intermediate states within the Riemann fan, which are then numerically calculated following the Rankine-Hugoniot (RH) jump relations across the eigenwaves or the spatial-temporal integral conservation \cite{Toro2009}.  The HLL scheme \cite{Harten_1983}, HLLC scheme \cite{Gurski2004,Li2005}, and HLLD scheme \cite{Miyoshi2005}, respectively assume that there are two, three, and five eigenwaves within the Riemann fan, leading to different levels of physical accuracy and numerical resolution. These methods have been widely implemented in multi-dimensional MHD codes \cite{Mignone_2007,Porth_2014,White_2016, Perri2022, Navarro_2025,Popovas2025}.

More specifically, while all of the HLL-type schemes mentioned above include two fast magnetoacoustic waves, the contact discontinuity has been included by the HLLC scheme, and the HLLD scheme further includes two Alfv\'en waves within the Riemann fan.
However, the slow magnetoacoustic mode has not been included in these HLL-type Riemann solvers, as the incompressible assumption used to compute the intermediate states does not allow the slow mode to exist within the Riemann fan. A solution is to extend the HLLEM scheme \cite{Einfeldt_1988}, by adding an anti-diffusive term to the two-wave HLL approximation \cite{DUMBSER2016}. This solution reverts to the HLL scheme when some of the eigenvalues of 
the hyperbolic system coincide, consequently having no complete set of linearly independent eigenvectors.

In this work, we propose a multi-state HLL-type scheme that includes the slow magnetoacoustic mode within the Riemann fan, thus providing in a seven-wave six-state Riemann solver. Specifically, we first briefly discuss the basics of the  space-time configuration  of HLL-type schemes in general. Then, in Section \ref{sec:compressibility}, we revise the flux conservation calculation used when solving the intermediate states,  allowing for the compressible assumption and providing a complete seven-wave approximation.  Subsequently, in Section \ref{sec:transverse}, we discuss the degenerate scenarios, providing approximate solutions accordingly, and then in Section \ref{sec:combine}
  we provide a simple strategy to smoothly combine the numerical approximations with and without degeneracy.  Numerical test cases are provided in Section \ref{sec:tests}, to showcase the performance of the present solution. Finally, we end this work with concluding remarks in the last section.

\section{Basics}
\subsection{The  HLL-type approximation}

For a piecewise-smooth weak solution of hyperbolic conservation laws, across each discontinuity, the solution should follow the RH jump relation 
\begin{equation} \label{eq:jump}
\mathbf{F}^{+}-\mathbf{F}^{-}=S\left(\mathbf{U}^{+}-\mathbf{U}^{-}\right),
\end{equation}
\noindent where $S$ is the propagation speed of the discontinuity,  $\mathbf{U}^{{\pm}}$ are the left and right states relative to the discontinuity, and $\mathbf{F}^{\pm}=\mathbf{F}(\mathbf{U}^{\pm})$.

Given arbitrary initial $\mathbf{U}^{\text{l,r}}$ which are the left and right states  that constitute a Riemann problem at a discretized interface, $\mathbf{U}^{\text{l,r}}$ may evolve and develop multiple self-similar discontinuities or eigenwaves in the $x-t$ space, across each of which a RH jump relation should be followed. 
It is easy to find that if we have $n$ eigenmodes from left to right of the Riemann fan, we may write the  conservation law as follows
\begin{equation} \label{eq:conservation}
\mathbf{F}^{\text{r}}-\mathbf{F}^{\text{l}}=\sum_{j=1}^{n} S_j(\mathbf{U}_{j}-\mathbf{U}_{j-1}),
\end{equation}
\noindent where $S_{j}$ is the propagation speed of the $j$-th eigenwave, and $\mathbf{U}_j$ is the intermediate state between the $j$-th and $j+1$-th eigenwaves, while $\mathbf{U}_0=\mathbf{U}^{\text{l}}$ and $\mathbf{U}_{n}=\mathbf{U}^{\text{r}}$. In particular, the HLL average 
\begin{equation} \label{eq:U_HLL}
\mathbf{U}^{\text{hll}}  =\frac{S^{\text{r}}_{\text{f}}\mathbf{U}^{\text{r}}-S^{\text{l}}_{\text{f}}\mathbf{U}^{\text{l}}-\mathbf{F}^{\text{r}}+\mathbf{F}^{\text{l}}}{S^{\text{r}}_{\text{f}}-S^{\text{l}}_{\text{f}}},
 \end{equation}
\noindent is the simplest example of Eq.~\eqref{eq:conservation} when there are only two discontinuities, which are  two fast magnetoacoustic waves for ideal MHD, throughout the Riemann fan.  For charity, hereafter, superscripts denote variable locations or the numerical methods used to define them, and subscripts indicate their physical relevance or coordinate direction. For example, $S^{\text{l,r}}_{\text{f}}$ denote the speeds of the left and right going fast magnetoacoustic waves.
 
 A multi-state HLL-type approach that includes all seven eigenmodes of the ideal MHD Riemann problem, hereafter denoted as HLLx (HLL-extended), can be written in a manner similar to other multi-state HLL-type schemes, to  provide the numerical fluxes  on the right-hand side of Eq.~\eqref{eq:semi-discrete}: 
\begin{eqnarray}  \label{eq:HLLD}
\mathbf{F}_{1/2}^{\text{hllx}} =
\left\{\begin{array}{ll}
\mathbf{F}^{\text{l}},&\text{if}~S^{\text{l}}_{\text{f}}>0,\\
\mathbf{F}^{\text{l}}+S^{\text{l}}_{\text{f}}(\mathbf{U}^{\text{l}}_{\text{f}}-\mathbf{U}^{\text{l}}),&\text{if}~S^{\text{l}}_{\text{f}}\le 0< S^{\text{l}}_{\text{a}}, \\ 
\mathbf{F}^{\text{l}}+S^{\text{l}}_{\text{f}}(\mathbf{U}^{\text{l}}_{\text{f}}-\mathbf{U}^{\text{l}})+S^{\text{l}}_{\text{a}}(\mathbf{U}^{\text{l}}_{\text{a}}-\mathbf{U}^{\text{l}}_{\text{f}}),&\text{if}~S^{\text{l}}_{\text{a}}\le 0< S^{\text{l}}_{\text{s}}, \\ 
\mathbf{F}^{\text{l}}+S^{\text{l}}_{\text{f}}(\mathbf{U}^{\text{l}}_{\text{f}}-\mathbf{U}^{\text{l}})+S^{\text{l}}_{\text{a}}(\mathbf{U}^{\text{l}}_{\text{a}}-\mathbf{U}^{\text{l}}_{\text{f}})+S^{\text{l}}_{\text{s}}(\mathbf{U}_{\text{s}}^{\text{l}}-\mathbf{U}^{\text{l}}_{\text{a}}),&\text{if}~S^{\text{l}}_{\text{s}}\le 0< S^{\text{m}}, \\ 
\mathbf{F}^{\text{r}}+S^{\text{r}}_{\text{f}}(\mathbf{U}^{\text{r}}_{\text{f}}-\mathbf{U}^{\text{r}})+S^{\text{r}}_{\text{a}}(\mathbf{U}^{\text{r}}_{\text{a}}-\mathbf{U}^{\text{r}}_{\text{f}})+S^{\text{r}}_{\text{s}}(\mathbf{U}^{\text{r}}_{\text{s}}-\mathbf{U}^{\text{r}}_{\text{a}}),&\text{if}~S^{\text{m}}\le 0 < S^{\text{r}}_{\text{s}}, \\
\mathbf{F}^{\text{r}}+S^{\text{r}}_{\text{f}}(\mathbf{U}^{\text{r}}_{\text{f}}-\mathbf{U}^{\text{r}})+S^{\text{r}}_{\text{a}}(\mathbf{U}^{\text{r}}_{\text{a}}-\mathbf{U}^{\text{r}}_{\text{f}}),&\text{if}~S^{\text{r}}_{\text{s}}\le 0 < S^{\text{r}}_{\text{a}}, \\
\mathbf{F}^{\text{r}}+S^{\text{r}}_{\text{f}}(\mathbf{U}^{\text{r}}_{\text{f}}-\mathbf{U}^{\text{r}}),&\text{if}~S^{\text{r}}_{\text{a}}\le 0< S^{\text{r}}_{\text{f}}, \\
\mathbf{F}^{\text{r}},&\text{if}~S^{\text{r}}_{\text{f}}\le 0.
\end{array}    \right.
\end{eqnarray}
\noindent where the subscripts f, a, and s, respectively, denote the propagation speeds of, or the intermediate states behind fast magnetoacoustic waves, Alfv\'en waves, and slow magnetoacoustic waves,  and the superscripts l, m, and r, respectively, denote the left, middle, and right states or speeds. The left and right unperturbed states $\mathbf{U}^{\text{l,r}}$, i.e., states outside the Riemann fan or   ahead of the fast waves, are the conservative variables on, for example, the left and right sides of an interface between two FV cells. The wave speeds and intermediate states are to be computed.  Note that the solutions of the wave speeds are not unique and may affect robustness \cite{EINFELDT1991}. 
It is straightforward to reduce Eq.~\eqref{eq:HLLD} for the HLL, HLLC, and HLLD schemes, and thus the details are omitted here. 

\subsection{Degenerated MHD modes}

Before designing the approximate Riemann solution, we discuss wave-mode degeneration and the resulting jump relations. 
When $B_x=0$, the seven-wave configuration degenerates to the three-wave configuration that includes two fast magnetoacoustic waves and a contact discontinuity, and an HLLC-type scheme suffices to capture all three existing eigenmodes. However, other scenarios exist, which were not usually investigated when designing an approximate Riemann solution. A detailed theoretical explanation can be found in, for example, Refs.~\cite{Goedbloed2004,Priest2014}, and a more systematic overview on this issue can be found in Refs. \cite{Isaacson_1992,Roe_1996}. Here we briefly discuss the properties of the Riemann problem in the degenerate scenarios. 

First, we examine the eigenwave speeds. The fast and slow magnetoacoustic speeds are 
  \begin{equation} \label{eq:sounds}
c_{\text{f,s}} =\sqrt{\frac{\gamma p_{\text{th}} + 2p_{\text{mag}}\pm\sqrt{\left(\gamma p_{\text{th}}+ 2p_{\text{mag}}\right)^2-4\gamma p_{\text{th}}B_x^2}}{2\rho}},
 \end{equation}
\noindent where $p_{\text{th}}
$  is plasma thermal pressure, $p_{\text{mag}}=\frac{1}{2}\mathbf{B}^2$ is magnetic pressure, and "$+$" and "$-$" in "$\pm$" respectively correspond to  the  f(ast) and s(low) speeds. Considering that the Alfv\'en speed is 
 \begin{equation}
c_{\text{a}}=\frac{|B_x|}{\sqrt{\rho}}
 \end{equation}
 \noindent the magnetoacoustic speeds may become equal  to the Alfv\'en speed or sound speed $\sqrt{{\gamma p_{\text{th}}}/{\rho}}$ when $|\mathbf{B}|=|B_x|$, as  Eq.~(\ref{eq:sounds}) becomes 
\begin{equation}  
c_{\text{f,s}} =\sqrt{\frac{\gamma p_{\text{th}} + 2p_{\text{mag}}\pm{|\gamma p_{\text{th}}- 2p_{\text{mag}}|}}{2\rho}}.
 \end{equation}

More specifically, when the Alfv\'en speed $c_{\text{a}}$ is faster than the sound speed, we have 
 \begin{eqnarray}   
  \left\{\begin{array}{c}
c_{\text{f}}=\frac{|B_x|}{\sqrt{\rho}},  \\
c_{\text{s}}=\sqrt{{\gamma p_{\text{th}}}/{\rho}}, 
 \end{array}    \right.
 \end{eqnarray}
 \noindent in which case the fast wave and the Alfv\'en wave merge. In contrary, when the Alfv\'en speed   is slower than the sound speed, we have 
 \begin{eqnarray}   
  \left\{\begin{array}{c}
c_{\text{f}}=\sqrt{{\gamma p_{\text{th}}}/{\rho}},  \\
c_{\text{s}}=\frac{|B_x|}{\sqrt{\rho}},
 \end{array}    \right.
 \end{eqnarray}
 \noindent in which case the slow wave and the Alfv\'en wave merge. The last special case is when the Alfv\'en speed and the sound speed are equal; consequently, all three eigenmodes degenerate into one mode.

 In the degenerate scenarios, what changes is not only the calculation of eigenwave speeds. More importantly, the jump relations also change/degenerate. 
We may have a linear analysis, assuming that across an eigenmode whose eigenspeed is $S=u\pm c_{\text{x}}$, where the subscript x indicates that the mode may be any one of these three modes, the variables change from $\rho$, $u$, and $p_{\text{th}}$ to $\rho+\text{d}\rho$, $u+\text{d}u$, and $p_{\text{th}}+\text{d}p_{\text{th}}$, while $B_x$ is naturally constant. Then, based on the jump relations in Eq.~\eqref{eq:jump} of the continuity equation and the longitudinal momentum equation across the given eigenwave, we have 
 \begin{eqnarray}   
  \left\{\begin{array}{c}
 \left(\rho+\text{d}\rho)(u+\text{d}u\right)-\rho u=S\left(\left(\rho+\text{d}\rho\right)-\rho\right),  \\
 \left(\rho+\text{d}\rho)(u+\text{d}u\right)^2+(p_{\text{th}}+\text{d}p_{\text{th}})-\rho u^2-p_{\text{th}}=S\left((\rho+\text{d}\rho)(u+\text{d}u)-\rho u\right).
 \end{array}\right.
 \end{eqnarray}
\noindent Without listing the derivation and keeping only the first-order perturbations, we obtain 
\begin{equation}
(S-u)\rho\text{d}u=\text{d}p_{\text{th}},
\end{equation} 
\noindent  in which $\text{d}u\ne 0$ can only be true when $S=u\pm \sqrt{{\gamma p_{\text{th}}}/{\rho}}$. Therefore, when the Alfv\'en mode merges with one of the magnetoacoustic modes, this degenerate mode is incompressible. 
If all three modes merge/degenerate into one mode, then this mode is still compressible. Thus, we must properly account for the compressibility of degenerate modes when designing the numerical solution, as the jump relations must be handled correctly.

\section{The seven-wave configuration} \label{sec:compressibility}

In this section, we discuss the solution when seven eigenmodes fully develop, without degeneracy.
\subsection{The magnetoacoustic speeds}
Ideal MHD problems may develop two types of compressible eigenmodes: the fast magnetoacoustic mode and the slow magnetoacoustic mode. As mentioned, the former is included in all HLL-type approximate Riemann solutions. 
Straightforwardly, we may calculate their propagation speeds as
 \begin{eqnarray}   \label{eq:SLSR} 
  \left\{\begin{array}{c}
S^{\text{l}}_{\text{f}}=\min\left(u^{\text{l}}, u^{\text{r}}\right)-\max\left(c_{\text{f}}^{\text{l}},c_{\text{f}}^{\text{r}}\right),  \\
S^{\text{r}}_{\text{f}}=\max\left(u^{\text{l}}, u^{\text{r}}\right)+\max\left(c_{\text{f}}^{\text{l}},c_{\text{f}}^{\text{r}}\right), 
 \end{array}    \right. 
 \end{eqnarray}
\noindent where $c_{\text{f}}^{\text{l,r}}$ are respectively the fast magnetoacoustic wave speeds calculated using the left and right unperturbed states $\mathbf{U}^{\text{l,r}}$. 
However, as mentioned, HLL-type schemes typically assume a constant longitudinal velocity component across the entire Riemann fan \cite{Toro1994,Batten_1997,Miyoshi2005}, excluding the possibility of including the slow magnetoacoustic mode. 

Nonetheless, the major restriction is that a multi-state HLL-type scheme calculates the eigenspeeds using the intermediate states between the Riemann fan. For example, the classic HLLD scheme calculates the Alfv\'en speeds using intermediate densities, which depend nonlinearly on the unperturbed states and thus cannot be provided in advance.
In this way, calculating the slow magnetoacoustic speed would be even more complicated, as it requires more variables. Therefore, we need a simplified solution, which we discuss in detail in Section \ref{sec:combine}. Here, the left- and right-going magnetoacoustic speeds are written as simple functions of left and right unperturbed states: 
\begin{equation}  \label{eq:SSLSSR}
S^{\text{l,r}}_{\text{s}}=S^{\text{l,r}}_{\text{s}}\left(\mathbf{U}^{\text{l}},\mathbf{U}^{\text{r}}\right),
\end{equation}
 \noindent which are not dependent on the estimates of the intermediate states.

\subsection{The conservation of the longitudinal momentum}

When the medium within the Riemann fan is compressible, we need to estimate the variation of the longitudinal velocity.
Density and the longitudinal velocity do not change across an Alfv\'en wave, and the longitudinal velocity does not change across a contact discontinuity. Therefore, to obtain the variations of density and the longitudinal velocity within the Riemann fan when including all eigenwaves, we need to solve seven
unknown variables. However, without knowing the transverse components of the magnetic field between fast and slow waves, we cannot provide the corresponding intermediate total pressures needed to solve the longitudinal-momentum jump relations. 

Across two fast waves and two slow waves, we can provide four jump relations for density. 
Then, as the jump relation of the longitudinal
momentum is not yet solvable, 
we rewrite Eq.~(\ref{eq:conservation}) to 
\begin{eqnarray}    
\mathbf{F}^{\text{l}}_{\text{s}}-\mathbf{F}^{\text{l}}=S^{\text{l}}_{\text{f}}(\mathbf{U}^{\text{l}}_{\text{f}}-\mathbf{U}^{\text{l}})+S^{\text{l}}_{\text{s}}(\mathbf{U}^{\text{l}}_{\text{s}}-\mathbf{U}^{\text{l}}_{\text{f}}), \label{eq:two_relations} \\ 
\mathbf{F}^{\text{r}}_{\text{f}}-\mathbf{F}^{\text{l}}_{\text{f}}=S^{\text{l}}_{\text{s}}(\mathbf{U}^{\text{l}}_{\text{s}}-\mathbf{U}^{\text{l}}_{\text{f}})+S^{\text{r}}_{\text{s}}(\mathbf{U}^{\text{r}}_{\text{f}}-\mathbf{U}^{\text{r}}_{\text{s}}), \label{eq:two_relations2} \\
\mathbf{F}^{\text{r}}-\mathbf{F}^{\text{r}}_{\text{s}}=S^{\text{r}}_{\text{s}}(\mathbf{U}^{\text{r}}_{\text{f}}-\mathbf{U}^{\text{r}}_{\text{s}})+S^{\text{r}}_{\text{f}}(\mathbf{U}^{\text{r}}-\mathbf{U}^{\text{r}}_{\text{f}}),  \label{eq:two_relations3} 
 \end{eqnarray}
\noindent for the longitudinal
momentum from left to right within the Riemann fan, separated by the compressible wave modes.   Then, we introduce two estimates that will be separately provided in Section \ref{sec:combine}, namely, the longitudinal velocity between the slow waves, denoted as $u^{\text{m(x)}}$, and the total pressure between the slow waves, denoted as $P^{\text{m(x)}}$. The total pressure is expected to be constant across the contact discontinuity. Having $u^{\text{m(x)}}$ ready separately removes one unknown, and being able to estimate $P^{\text{m(x)}}$ means that we may use two relations,  Eq.~\eqref{eq:two_relations} and Eq.~\eqref{eq:two_relations3}, consequently making the problem well-posed by solving six unknowns with six equations. Moreover, the flexibility of choosing $u^{\text{m(x)}}$ and $P^{\text{m(x)}}$ is necessary to deal with the degenerated scenarios.

Having the four jump relations and two conservation relations mentioned above, we first solve the  intermediate density and longitudinal velocity between the left-going fast and slow magnetoacoustic waves, using two jump relations and one conservation relation:
\begin{eqnarray}    
\rho^{\text{l}}_{\text{f}}=\rho^{\text{l}}\frac{S^{\text{l}}_{\text{f}}-u^{\text{l}}}{S^{\text{l}}_{\text{f}}-u^{\text{l}}_{\text{f}}}, \label{eq:left_relations} \\ 
\rho^{\text{l}}_{\text{f}}=\rho_{\text{s}}^{\text{l}}\frac{S^{\text{l}}_{\text{s}}-u^{\text{m(x)}}}{S^{\text{l}}_{\text{s}}-u^{\text{l}}_{\text{f}}}, \label{eq:left_relations2} \\
u^{\text{l}}_{\text{f}}=\frac{\rho^{\text{l}}_{\text{s}} u^{\text{m(x)}}(S^{\text{l}}_{\text{s}}-u^{\text{m(x)}})-P^{\text{m(x)}}-\rho^{\text{l}} u^{\text{l}}(S_{\text{f}}^{\text{l}}-u^{\text{l}})+P^{\text{l}}}{\rho^{\text{l}}_{\text{s}}(S^{\text{l}}_{\text{s}}-u^{\text{m(x)}})-\rho^{\text{l}}(S^{\text{l}}_{\text{f}}-u^{\text{l}})},  \label{eq:left_relations3} 
 \end{eqnarray}
\noindent in which we need to solve $\rho^{\text{l}}_{\text{f}}$, $u^{\text{l}}_{\text{f}}$, and $\rho_{\text{s}}^{\text{l}}$.  
We then substitute Eq.~\eqref{eq:left_relations} into Eq.~\eqref{eq:left_relations2}   to obtain a relation between $\rho_{\text{s}}^{\text{l}}$ and the known variables ahead of the left-going fast wave. Then, replacing $\rho_{\text{s}}^{\text{l}}$ in Eq.~\eqref{eq:left_relations3}  with the obtained relation, we have 
\begin{equation} 
u^{\text{l}}_{\text{f}}=\frac{\rho^{\text{l}}u^{\text{m(x)}}(S^{\text{l}}_{\text{f}}-u^{\text{l}})(S^{\text{l}}_{\text{s}}-u^{\text{l}}_{\text{f}})-(S^{\text{l}}_{\text{f}}-u^{\text{l}}_{\text{f}})\left(P^{\text{m(x)}}+\rho^{\text{l}} u^{\text{l}}(S_{\text{f}}^{\text{l}}-u^{\text{l}})-P^{\text{l}}\right)}{\rho^{\text{l}}(S^{\text{l}}_{\text{f}}-u^{\text{l}})(S^{\text{l}}_{\text{s}}-S^{\text{l}}_{\text{f}})},
\end{equation} 
\noindent which leads to a unique solution 
\begin{equation} \label{eq:left_u}
u^{\text{l}}_{\text{f}}=\frac{\rho^{\text{l}}u^{\text{m(x)}}S^{\text{l}}_{\text{s}}(S^{\text{l}}_{\text{f}}-u^{\text{l}})-S^{\text{l}}_{\text{f}}\left(P^{\text{m(x)}}+\rho^{\text{l}} u^{\text{l}}(S_{\text{f}}^{\text{l}}-u^{\text{l}})-P^{\text{l}}\right)}{\rho^{\text{l}}(S^{\text{l}}_{\text{f}}-u^{\text{l}})(S^{\text{l}}_{\text{s}}-S^{\text{l}}_{\text{f}})+\rho^{\text{l}}u^{\text{m(x)}}(S^{\text{l}}_{\text{f}}-u^{\text{l}})-\left(P^{\text{m(x)}}+\rho^{\text{l}} u^{\text{l}}(S_{\text{f}}^{\text{l}}-u^{\text{l}})-P^{\text{l}}\right)}.
\end{equation} 
\noindent Without repeating a similar process, the longitudinal velocity  between the right-going fast and slow waves can be given as 
\begin{equation} \label{eq:right_u}
u^{\text{r}}_{\text{f}}=\frac{S^{\text{r}}_{\text{f}}\left(P^{\text{m(x)}}+\rho^{\text{r}} u^{\text{r}}(S_{\text{f}}^{\text{r}}-u^{\text{r}})-P^{\text{r}}\right)-\rho^{\text{r}}u^{\text{m(x)}}S^{\text{r}}_{\text{s}}(S^{\text{r}}_{\text{f}}-u^{\text{r}})}{\rho^{\text{r}}(S^{\text{r}}_{\text{f}}-u^{\text{r}})(S^{\text{r}}_{\text{f}}-S^{\text{r}}_{\text{s}})-\rho^{\text{r}}u^{\text{m(x)}}(S^{\text{r}}_{\text{f}}-u^{\text{r}})+\left(P^{\text{m(x)}}+\rho^{\text{r}} u^{\text{r}}(S_{\text{f}}^{\text{r}}-u^{\text{r}})-P^{\text{r}}\right)}.
\end{equation} 
\noindent Then, the intermediate densities $\rho^{\text{l,r}}_{\text{f,s}}$ can be calculated straightforwardly following their respective jump relations.

\subsection{The transverse components}

When $B_x\ne 0$, the transverse components of the intermediate velocity and magnetic field change across fast, Alfv\'en, and slow waves, but are assumed to be constant across the contact discontinuity. Therefore, we  now need to solve for 20 unknowns, 8 of which can be calculated following the jump relations across the fast waves, resulting in  
 \begin{eqnarray}  
  v_{\text{f}}^{\alpha}=v^{\alpha}-B_xB_{y}^{\alpha}\frac{u_{\text{f}}^{\alpha}-u^{\alpha}}{\rho^{\alpha}(S^{\alpha}_{\text{f}}-u^{\alpha})(S^{\alpha}_{\text{f}}-u_{\text{f}}^{\alpha})-B_x^2}, \label{eq:fast_transverse}  \\
  B_{y,\text{f}}^{\alpha}={B}_{y}^{\alpha}\frac{\rho^{\alpha}(S^{\alpha}_{\text{f}}-u^{\alpha})^2-B_x^2
 }{\rho^{\alpha}(S^{\alpha}_{\text{f}}-u^{\alpha})(S^{\alpha}_{\text{f}}-u_{\text{f}}^{\alpha})-B_{x}^2}, \label{eq:fast_transverse2}  
 \end{eqnarray}
\noindent where $\alpha=\text{l}$ or r.  It is also trivial to change the formulae for $w_{\text{f}}^{\alpha}$  and $B_{z,\text{f}}^{~\alpha}$, and thus the corresponding details are omitted for simplicity.

As mentioned in Ref.~\cite{Miyoshi2005}, the jump relations across the Alfv\'en waves are not solvable based on the assumptions used by the classic HLLD scheme. In fact, these jump relations provide only 4 independent constraints on the intermediate variables. Therefore, combining these 4 constraints with the 8 jump relations across the slow waves yields a well-posed problem for the 12 remaining unknowns. Specifically,   using the jump relations across the (left and right) slow waves, we obtain the following formulas
 \begin{eqnarray}    
B_{y,\text{a}}^{\alpha}=B_{y,\text{s}}\frac{\rho^{\alpha}_{\text{f}}(S^{\alpha}_{\text{s}}-u^{\alpha}_{\text{f}})(S^{\alpha}_{\text{s}}-u^{\text{m}})-B_{x}^2}{\rho^{\alpha}_{\text{f}}(S^{\alpha}_{\text{s}}-u^{\alpha}_{\text{f}})^2-B_x^2
 },  \label{eq:b_transverse} \\
B_{z,\text{a}}^{\alpha}=B_{z,\text{s}}\frac{\rho^{\alpha}_{\text{f}}(S^{\alpha}_{\text{s}}-u^{\alpha}_{\text{f}})(S^{\alpha}_{\text{s}}-u^{\text{m}})-B_{x}^2}{\rho^{\alpha}_{\text{f}}(S^{\alpha}_{\text{s}}-u^{\alpha}_{\text{f}})^2-B_x^2
 }, 
 \end{eqnarray}
 \noindent and 
  \begin{eqnarray}     
v^{\alpha}_{\text{a}}=v_{\text{s}}+B_xB_{y,\text{s}}\frac{u^{\text{m}}-u^{\alpha}_{\text{f}}}{\rho^{\alpha}_{\text{f}}(S^{\alpha}_{\text{s}}-u^{\alpha}_{\text{f}})^2-B_x^2
 }, \label{eq:v_trans} \\ 
w^{\alpha}_{\text{a}}=w_{\text{s}}+B_xB_{z,\text{s}}\frac{u^{\text{m}}-u^{\alpha}_{\text{f}}}{\rho^{\alpha}_{\text{f}}(S^{\alpha}_{\text{s}}-u^{\alpha}_{\text{f}})^2-B_x^2}.   \label{eq:v_trans2}
 \end{eqnarray}
\noindent  We have the Alfv\'en speeds written as:
  \begin{equation}     \label{eq:alfvenSpeed}
   \left\{\begin{array}{c}
S^{\text{l}}_{\text{a}}=u^{\text{l}}_{\text{f}}-\frac{|B_x|}{\sqrt{\rho_{\text{f}}^{\text{l}}}},  \\
S^{\text{r}}_{\text{a}}=u^{\text{r}}_{\text{f}}+\frac{|B_x|}{\sqrt{\rho_{\text{f}}^{\text{r}}}}. 
 \end{array}    \right.
 \end{equation}
 
\noindent Then, the jump relations across the (left and right) Alfv\'en waves provide 
\begin{eqnarray} 
v_{\text{a}}^{\alpha}=v_{\text{f}}^{\alpha}+\frac{B_x (B_{y,\text{f}}^{\alpha} - B_{y,\text{a}}^{\alpha})}
{\rho_{\text{f}}^{\alpha}(S_{\text{a}}^{\alpha}-u_{\text{f}}^{\alpha})}, \label{eq:v_alfven}\\
w_{\text{a}}^{\alpha}=w_{\text{f}}^{\alpha}+\frac{B_x (B_{z,\text{f}}^{\alpha} - B_{z,\text{a}}^{\alpha})}
{\rho_{\text{f}}^{\alpha}(S_{\text{a}}^{\alpha}-u_{\text{f}}^{\alpha})}. \label{eq:v_alfven2}
\end{eqnarray}

Using Eqs.~(\ref{eq:b_transverse}) $\sim$ (\ref{eq:v_alfven2}), we may derive the solutions below. To simplify the process, we first give the following notations:
 \begin{eqnarray}     
   \left\{\begin{array}{c}
A=\frac{\rho^{\text{l}}_{\text{f}}(S^{\text{l}}_{\text{s}}-u^{\text{l}}_{\text{f}})(S^{\text{l}}_{\text{s}}-u^{\text{m}})-B_{x}^2}{\rho^{\text{l}}_{\text{f}}(S^{\text{l}}_{\text{s}}-u^{\text{l}}_{\text{f}})^2-B_x^2
 },  \\
B=\frac{\rho^{\text{r}}_{\text{f}}(S^{\text{r}}_{\text{s}}-u^{\text{r}}_{\text{f}})(S^{\text{r}}_{\text{s}}-u^{\text{m}})-B_{x}^2}{\rho^{\text{r}}_{\text{f}}(S^{\text{r}}_{\text{s}}-u^{\text{r}}_{\text{f}})^2-B_x^2
 }, \\
C=B_x\frac{u^{\text{m}}-u^{\text{l}}_{\text{f}}}{\rho^{\text{l}}_{\text{f}}(S^{\text{l}}_{\text{s}}-u^{\text{l}}_{\text{f}})^2-B_x^2
 },\\
D=B_x\frac{u^{\text{m}}-u^{\text{r}}_{\text{f}}}{\rho^{\text{r}}_{\text{f}}(S^{\text{r}}_{\text{s}}-u^{\text{r}}_{\text{f}})^2-B_x^2
 },\\
E=\frac{B_x}
{\rho_{\text{f}}^{\text{l}}(S_{\text{a}}^{\text{l}}-u_{\text{f}}^{\text{l}})}=-\frac{\text{sign}(B_x)}{\sqrt{\rho_{\text{f}}^{\text{l}}}},\\
F=\frac{\text{sign}(B_x)}{\sqrt{\rho_{\text{f}}^{\text{r}}}}.
 \end{array}    \right.
 \end{eqnarray}
 Furthermore, given that the equations for components in the $y$- and $z$-directions are independent yet similar, we solve the $y$ components  $v$ and $B_y$, as an example. 
The equations are 
 \begin{eqnarray}     
   \left\{\begin{array}{c}
B_{y,\text{a}}^{\text{l}}-B_{y,\text{s}}A=0,  \\
B_{y,\text{a}}^{\text{r}}-B_{y,\text{s}}B=0, \\
v^{\text{l}}_{\text{a}}-v_{\text{s}}-B_{y,\text{s}}C=0,\\
v^{\text{r}}_{\text{a}}-v_{\text{s}}-B_{y,\text{s}}D=0,\\
v_{\text{a}}^{\text{l}}-v_{\text{f}}^{\text{l}}-E(B_{y,\text{f}}^{\text{l}} - B_{y,\text{a}}^{\text{l}})=0,\\
v_{\text{a}}^{\text{r}}-v_{\text{f}}^{\text{r}}-F(B_{y,\text{f}}^{\text{r}} - B_{y,\text{a}}^{\text{r}})=0,
 \end{array}    \right.
 \end{eqnarray}
 \noindent where all variables behind the fast waves have been solved by Eqs.~\eqref{eq:fast_transverse} -- \eqref{eq:fast_transverse2}. This system of linear equations is easy to solve, and thus the solutions between the slow waves are directly given as 
 \begin{eqnarray}     
B_{y,s}
=
\frac{
v_{\text{f}}^{\text{l}} + E B_{y,{\text{f}}}^{\text{l}} - v_{\text{f}}^{\text{r}} - F B_{y,\text{f}}^{\text{r}}
}{C-D+AE-BF
},\label{eq:ss}
\\
v_s
=\frac{(BF+D)(v_{\text{f}}^{\text{l}}+EB_{y,\text{f}}^{\text{l}})-(AE+C)(v_{\text{f}}^{\text{r}}+FB_{y,\text{f}}^{\text{r}})}{D-C+BF-AE},\label{eq:ss2}
 \end{eqnarray}
\noindent while the rest of the solution can be easily obtained by substituting Eq.~(\ref{eq:ss}) and Eq.~(\ref{eq:ss2}) into Eqs.~(\ref{eq:b_transverse}) -- (\ref{eq:v_alfven2}).
The $z$ components, $w$ and $B_z$, can be calculated using the same formulas by replacing corresponding $y$-components, without needing to change the coefficients.

\subsection{The energy equation} \label{sec:closure}

To calculate the intermediate internal energy, we cannot assume that it is constant over the whole Riemann fan. We reuse the rationale of the HLLD-ec scheme \cite{zhang2026}, which is not repeated here for simplicity, and the intermediate internal energies behind the fast waves are given as  
 \begin{equation} \label{eq:fast_e}
(\rho e)^{\alpha}_{\text{f}}=(\rho e)^{\alpha}\frac{S^{\alpha}- \gamma u^{\alpha}}{S^{\alpha}- \gamma u_{\text{f}}^{\alpha}},
\end{equation}
 \noindent which remain the same across the Alfv\'en waves.
  Then the intermediate internal energy between the slow waves is given using the HLL-average  
 \begin{equation} \label{eq:ee}
(\rho e)_{\text{s}}=\frac{\left(S^{\text{r}}_{\text{f}}- \gamma u^{\text{r}}\right)(\rho e)^{\text{r}}-\left(S^{\text{l}}_{\text{f}}-\gamma u^{\text{l}}\right)(\rho e)^{\text{l}}-\left(S^{\text{r}}_{\text{f}}-S^{\text{r}}_{\text{s}}\right)(\rho e)^{\text{r}}_{\text{f}}+\left(S^{\text{l}}_{\text{f}}-S^{\text{l}}_{\text{s}}\right)(\rho e)^{\text{l}}_{\text{f}}}{S^{\text{r}}_{\text{s}}-S^{\text{l}}_{\text{s}}}.
\end{equation}
\noindent  We note that internal energy is not a conservative variable, so using this formula does not ensure internal energy conservation. 

Finally, again following the energy consistency condition \cite{zhang2026}, the intermediate total energies are given as 
 \begin{eqnarray}     
E^{\alpha}_{\text{f}}=\left(\rho e\right)^{\alpha}_{\text{f}}+\frac{1}{2}\rho^{\alpha}_{\text{f}}(\mathbf{V}^{\alpha}_{\text{f}})^2+\frac{1}{2}(\mathbf{B}^{\alpha}_{\text{f}})^2, \\
E^{\alpha}_{\text{a}}=\left(\rho e\right)^{\alpha}_{\text{f}}+\frac{1}{2}\rho^{\alpha}_{\text{f}}(\mathbf{V}^{\alpha}_{\text{a}})^2+\frac{1}{2}(\mathbf{B}^{\alpha}_{\text{a}})^2, \\
E_{\text{s}}=\left(\rho e\right)_{\text{s}}+\frac{1}{2}\rho_{\text{s}}\mathbf{V}_{\text{s}}^2+\frac{1}{2}\mathbf{B}_{\text{s}}^2,
 \end{eqnarray}
 \noindent in which, across the Alfv\'en mode, no change appears in the internal energy. 
 
 It is worth noting that when no slow mode is included, the numerical diffusion of, for example, the classic HLLD Riemann solver may not be sufficient to stabilise numerical solutions, and extra diffusion may be needed \cite{zhang2026}.
The present solution thus naturally imposes appropriate numerical diffusion, particularly when the slow mode is dominant.

\section{Degenerated scenarios} \label{sec:transverse}

We note that the jump relations used in the last section assume that all seven eigenmodes develop completely. When the eigenmodes degenerate, some relations may need revision. Moreover, solutions for different scenarios must be properly integrated without singularities. Therefore, we discuss the degenerate scenarios in this section. 

\subsection{$B_x=0$}

When $B_x=0$, the seven-wave configuration within the Riemann fan reduces to a three-wave configuration.  Therefore, we only need an HLLC-type scheme, and  $u^{\text{m}}$ has to be constant over the entire Riemann fan. In this scenario, we use the classic definition used by HLL-type schemes, i.e.,
\begin{equation} \label{eq:SM}
u^{\text{m(hll)}}=\frac{\rho^{\text{r}} u^{\text{r}}(S_{\text{f}}^{\text{r}}-u^{\text{r}})-P^{\text{r}}-\rho^{\text{l}} u^{\text{l}}(S_{\text{f}}^{\text{l}}-u^{\text{l}})+P^{\text{l}}}{\rho^{\text{r}}(S^{\text{r}}_{\text{f}}-u^{\text{r}})-\rho^{\text{l}}(S^{\text{l}}_{\text{f}}-u^{\text{l}})},
\end{equation}
\noindent  ensuring the conservation of the longitudinal momentum.

Moreover, when $B_x\rightarrow 0$, the slow magnetoacoustic speeds and Alfv\'en speeds also  reach zero, i.e., $c_{\text{s}}\rightarrow 0$ and  $c_{\text{a}}\rightarrow 0$.  Then the jump relations across the fast wave can also be simplified, and thus Eqs.~(\ref{eq:left_u}) and (\ref{eq:right_u})  reduce to 
\begin{equation}    \label{eq:degeneratedUm}
  u^{\text{l}}_{\text{f}}=\frac{\rho^{\text{l}}u^{\text{m}}(S^{\text{l}}_{\text{f}}-u^{\text{l}})(S^{\text{l}}_{\text{s}}-S^{\text{l}}_{\text{f}})}{\rho^{\text{l}}(S^{\text{l}}_{\text{f}}-u^{\text{l}})(S^{\text{l}}_{\text{s}}-S^{\text{l}}_{\text{f}})}=u^{\text{m}}, 
   \end{equation}
\noindent and 
   \begin{equation}
    u^{\text{r}}_{\text{f}}=u^{\text{m}}, 
 \end{equation}
\noindent  if the total pressure $P^{\text{m(x)}}$ is calculated as
\begin{equation} \label{eq:Pmx}
P^{\text{m(hll)}}= \frac{\rho^{\text{r}}(S_{\text{f}}^{\text{r}}-u^{\text{r}})P^{\text{l}}-\rho^{\text{l}}(S_{\text{f}}^{\text{l}}-u^{\text{l}})P^{\text{r}}+\rho^{\text{l}}(S_{\text{f}}^{\text{l}}-u^{\text{l}})\rho^{\text{r}}(S_{\text{f}}^{\text{r}}-u^{\text{r}})(u^{\text{r}}-u^{\text{l}})}{\rho^{\text{r}}(S_{\text{f}}^{\text{r}}-u^{\text{r}})-\rho^{\text{l}}(S^{\text{l}}_{\text{f}}-u^{\text{l}})}.
\end{equation}

Correspondingly, the transverse components of both the velocity and magnetic fields directly behind the fast waves can be calculated without revising the formulas in the last section.
All the other intermediate states needed can be calculated using the HLLD-ec scheme \cite{zhang2026}.
Nonetheless, as $u^{\text{m(hll)}}$ is not used as a general solution of $u^{\text{m}}$, a basic principle needed here is that when $B_x\rightarrow 0$, we should have $u^{\text{m}}\rightarrow u^{\text{m(hll)}}$. 


\subsection{$B_x\ne 0$}

When $B_x\ne 0$, the solution may still degenerate into a three-wave configuration, which, however, is different from the scenario when $B_x= 0$.
We consider several scenarios below. While the left- and right-going waves can, in theory, fall into different scenarios, we consider both sides in the same scenario for simplicity. 

\subsubsection{ $c_{\text{s}}\approx c_{\text{a}}$, and $c_{\text{f}}> c_{\text{a}}$}

As explained, in this scenario the Alfv\'en mode merges with the slow mode, resulting in an incompressible mode, while the fast mode remains compressible.  
In this case, the density between the fast waves only changes when crossing the contact discontinuity, and the longitudinal velocity should be constant within the Riemann fan. Therefore,
we may reuse Eq.~\eqref{eq:SM} and Eq.~\eqref{eq:Pmx}, and the  jump relations behind the fast waves.

However, as the Alfv\'en modes may still exist, we need to calculate the intermediate states between the Alfv\'en modes. As has been explained in the last section, the jump relations across two Alfv\'en modes can only provide 4 independent constraints, and thus, to be well-posed, the transverse components of the velocity and magnetic fields need to be constant between the left- and right-going Alfv\'en modes. This is also a physical assumption since the total pressure should not change across the contact discontinuity. Solving Eq.~\eqref{eq:v_alfven} and Eq.~\eqref{eq:v_alfven2} for the left and right sides at the same time, we obtain 
\begin{eqnarray}    
v_{\text{a}}=\frac{\sqrt{\varrho^{\text{l}}_{\text{f}}}v_{\text{f}}^{\text{l}}+\sqrt{\varrho^{\text{r}}_{\text{f}}}v^{\text{r}}_{\text{f}}+({B}_{y,\text{f}}^{\text{r}}- {B}_{y,\text{f}}^{\text{l}})\text{sign}(B_{x})}{\sqrt{\varrho^{\text{l}}_{\text{f}}}+\sqrt{\varrho^{\text{r}}_{\text{f}}}}, \label{eq:degenerateHLLD}\\ 
{B}_{y,\text{a}}=\frac{\sqrt{\varrho^{\text{r}}_{\text{f}}}{B}_{y,\text{f}}^{\text{l}}+\sqrt{\varrho^{\text{l}}_{\text{f}}}{B}_{y,\text{f}}^{\text{r}}+\sqrt{\varrho^{\text{l}}_{\text{f}}\varrho^{\text{r}}_{\text{f}}}( v_{\text{f}}^{\text{r}}- v_{\text{f}}^{\text{l}})\text{sign}(B_{x})}{\sqrt{\varrho^{\text{l}}_{\text{f}}}+\sqrt{\varrho^{\text{r}}_{\text{f}}}}. \label{eq:degenerateHLLD2}
 \end{eqnarray}
 \noindent The components along the $z$-direction are similar and thus omitted here. In principle, the seven-wave configuration should degenerate to the present solution asymptotically when $ c_{\text{s}}\rightarrow c_{\text{a}}$, and thus we should have $u^{\text{m}}\rightarrow u^{\text{m(hll)}}$ at the same time. 
 
 We can also observe that Eq.~\eqref{eq:degenerateHLLD} and Eq.~\eqref{eq:degenerateHLLD2} are, in fact, the ones used by the classic HLLD scheme \cite{Miyoshi2005}. Therefore, the classic HLLD scheme actually assumes a specific space-time wave configuration of the present degenerate scenario. Since the HLLD scheme was designed for this scenario, we can reuse the HLLD-ec formulas \cite{zhang2026} for it. However, HLLD-type schemes use the five-wave configuration to cover all scenarios, including those in which the slow mode is dominant. In contrast, we separately consider scenarios in which the slow mode is compressible or degenerates. 
 
The HLLD-ec scheme abandons the constant total pressure assumption within the Riemann fan, but this is not necessary in the present scenario, where we may expect the total pressure to be constant within the Riemann fan. Nonetheless, we do not explicitly impose the constant total pressure constraint. Instead, when the jump relations across two Alfv\'en waves are properly addressed, the constraint should be satisfied automatically.

 \subsubsection{$c_{\text{f}}\approx c_{\text{a}}$, and $c_{\text{s}}<c_{\text{a}}$} \label{sec:important}

In this scenario, the fast magnetoacoustic speed and the Alfv\'en speed become close, and this degenerate mode is incompressible. We thus have $u^{\text{l,r}}_{\text{f,a}}\rightarrow u^{\text{l,r}}$ and $\rho^{\text{l,r}}_{\text{f,a}}\rightarrow \rho^{\text{l,r}}$. 

As in the general seven-wave scenario, the difficulty in providing a solution is that the intermediate longitudinal velocity between the slow waves cannot be calculated directly using the HLL average, because the transverse magnetic field components are needed in the numerical fluxes behind the fast-Alfv\'en compound waves, but these components are unknown. The only exception is when the amplitude of the fast-Alfv\'en compound mode is zero; then we may use 
\begin{equation} \label{eq:SM_slow}
u^{\text{m(hll(s))}}=\frac{\rho^{\text{r}} u^{\text{r}}(S_{\text{s}}^{\text{r}}-u^{\text{r}})-P^{\text{r}}-\rho^{\text{l}} u^{\text{l}}(S_{\text{s}}^{\text{l}}-u^{\text{l}})+P^{\text{l}}}{\rho^{\text{r}}(S^{\text{r}}_{\text{s}}-u^{\text{r}})-\rho^{\text{l}}(S^{\text{l}}_{\text{s}}-u^{\text{l}})},
\end{equation}
 \noindent and 
 \begin{equation} \label{eq:Pmx_slow}
P^{\text{m(hll(s))}}= \frac{\rho^{\text{r}}(S_{\text{s}}^{\text{r}}-u^{\text{r}})P^{\text{l}}-\rho^{\text{l}}(S_{\text{s}}^{\text{l}}-u^{\text{l}})P^{\text{r}}+\rho^{\text{l}}(S_{\text{s}}^{\text{l}}-u^{\text{l}})\rho^{\text{r}}(S_{\text{s}}^{\text{r}}-u^{\text{r}})(u^{\text{r}}-u^{\text{l}})}{\rho^{\text{r}}(S_{\text{s}}^{\text{r}}-u^{\text{r}})-\rho^{\text{l}}(S^{\text{l}}_{\text{s}}-u^{\text{l}})},
\end{equation}
\noindent where the propagation speeds of fast waves originally used by the HLL-average are replaced by those of the slow waves, the fastest compressible waves in this scenario. In Section \ref{sec:combine}, we will discuss the generalisation of the solutions.

As $c_{\text{f}}\rightarrow c_{\text{a}}$, we have $u_{\text{f}}^{\alpha}\rightarrow u^{\alpha}$.
Consequently,  both Eq.~\eqref{eq:fast_transverse} and Eq.~\eqref{eq:fast_transverse2} degenerate to 
 \begin{eqnarray}  
  v_{\text{f}}^{\alpha}=v^{\alpha}, \label{eq:vf_degen}\\
  B_{y,\text{f}}^{\alpha}={B}_{y}^{\alpha}.  \label{eq:byf_degen}
 \end{eqnarray}
\noindent Then the seven-wave configuration reduces to the present scenario, in theory, without needing to revise the formulas.  However, in practice we may use the exact formulas to avoid small numerical residuals affecting the result.

\subsubsection{$c_{\text{f}}\approx c_{\text{s}} \approx c_{\text{a}}$}

We have explained that, in this scenario, the degenerate mode is compressible, and the only mode between them is the contact discontinuity. Therefore, we may reuse Eq.~\eqref{eq:SM} and Eq.~\eqref{eq:Pmx}, and the densities can be calculated following the jump relations across the fast waves.

Additionally, to avoid numerical singularity, we directly use   
\begin{equation}
   u_{\text{f}}^{\alpha}=u_{\text{s}}^{\alpha}=u^{\text{m(hll)}},
\end{equation}
\noindent and the same degeneracy leading to Eqs.~\eqref{eq:vf_degen}--\eqref{eq:byf_degen},  while the original seven-wave solutions may become indeterminate. Then, formulas 
 Eqs.~\eqref{eq:degenerateHLLD}--\eqref{eq:degenerateHLLD2}, and Eqs.~\eqref{eq:fast_e}--\eqref{eq:ee} can be reused to calculate the other intermediate states.  

Apparently, since the three wave modes merge, we may again assume that the total pressure is constant within the Riemann fan.

\section{Closure} \label{sec:combine}

We have discussed the degeneracy of eigenmodes in the last section.  In general, if the fast mode is the only compressible eigenmode or all nonlinear modes overlap, we may reuse Eq.~\eqref{eq:SM} and Eq.~\eqref{eq:Pmx} to calculate the longitudinal velocity and the total pressure within the Riemann fan. When the slow mode is the only compressible mode, and meanwhile the amplitude of the Alfv\'en mode is zero, the HLL-average can again be used, except that the propagation speeds of the fast waves should be replaced by those of the slow waves.

To estimate the intermediate states between two slow waves when all fast and slow waves in the Riemann fan are compressible, we assume that the total compressibility of two waves can be represented by an "equivalent" compressible wave. Then, we may use two "equivalent" compressible waves, one left-going and one right-going, to constrain (I) the longitudinal velocity and (II) the total pressure between them. We know that having two waves is sufficient for the solutions of these two variables to be well-posed.

Apparently, we expect the propagation speed of the left- or right-going "equivalent" wave to be a function of the propagation speeds of the left- or right-going fast and slow waves. More importantly, given that the compressibility of a fast or slow wave disappears when it is overlapped with an Alfv\'en wave, we may adjust the function according to the Alfv\'en speed. Considering that the square of an eigenspeed is proportional to density, which is directly related to compressibility, we  use 
\begin{equation}
 c_{\text{e}}^{\alpha} =  \frac{\left(({c}^{\alpha}_{\text{f}})^2-({c}^{\alpha}_{\text{a}})^2\right) c_{\text{f}}^{\alpha}+ \left(({c}^{\alpha}_{\text{a}})^2-({c}^{\alpha}_{\text{s}})^2\right)c_{\text{s}}^{\alpha}}{({c}^{\alpha}_{\text{f}})^2-({c}^{\alpha}_{\text{s}})^2},
\end{equation}
\noindent where the eigenspeeds are calculated using left or right unperturbed states. Then the propagation speeds of the "equivalent" waves are
 \begin{eqnarray}   \label{eq:SeLSeR} 
  \left\{\begin{array}{c}
S^{\text{l}}_{\text{e}}=\min\left(u^{\text{l}}, u^{\text{r}}\right)-\max\left(c_{\text{e}}^{\text{l}},c_{\text{e}}^{\text{r}}\right),  \\
S^{\text{r}}_{\text{e}}=\max\left(u^{\text{l}}, u^{\text{r}}\right)+\max\left(c_{\text{e}}^{\text{l}},c_{\text{e}}^{\text{r}}\right).
 \end{array}    \right. 
 \end{eqnarray}

Now, the  HLL-average  can be used to provide the two remaining unknown intermediate variables:
 \begin{equation}  \label{eq:um}
u^{\text{m}}=\frac{\rho^{\text{r}} u^{\text{r}}(S_{\text{e}}^{\text{r}}-u^{\text{r}})-P^{\text{r}}-\rho^{\text{l}} u^{\text{l}}(S_{\text{e}}^{\text{l}}-u^{\text{l}})+P^{\text{l}}}{\rho^{\text{r}}(S^{\text{r}}_{\text{e}}-u^{\text{r}})-\rho^{\text{l}}(S^{\text{l}}_{\text{e}}-u^{\text{l}})},
\end{equation}
 \noindent and 
 \begin{equation} \label{eq:Pm}
P^{\text{m}}= \frac{\rho^{\text{r}}(S_{\text{e}}^{\text{r}}-u^{\text{r}})P^{\text{l}}-\rho^{\text{l}}(S_{\text{e}}^{\text{l}}-u^{\text{l}})P^{\text{r}}+\rho^{\text{l}}(S_{\text{e}}^{\text{l}}-u^{\text{l}})\rho^{\text{r}}(S_{\text{e}}^{\text{r}}-u^{\text{r}})(u^{\text{r}}-u^{\text{l}})}{\rho^{\text{r}}(S_{\text{e}}^{\text{r}}-u^{\text{r}})-\rho^{\text{l}}(S^{\text{l}}_{\text{e}}-u^{\text{l}})}.
\end{equation}
\noindent 
It is easy to see that Eqs.~\eqref{eq:um} and \eqref{eq:Pm} exactly recover the degenerate scenarios. The numerical solution may be more robust when $S^{\text{l,r}}_{\text{e}}$ are closer to $S^{\text{l,r}}_{\text{f}}$  than they should be. In the next section, we also numerically examine their performance in general scenarios. 

Finally, we provide the last missing component 
 \begin{eqnarray}   \label{eq:SSLSSR_full} 
  \left\{\begin{array}{c} 
S^{\text{l}}_{\text{s}}=\min\left(u^{\text{l}}, u^{\text{r}}\right)-\max\left(c_{\text{s}}^{\text{l}},c_{\text{s}}^{\text{r}}\right),  \\
S^{\text{r}}_{\text{s}}=\max\left(u^{\text{l}}, u^{\text{r}}\right)+\max\left(c_{\text{s}}^{\text{l}},c_{\text{s}}^{\text{r}}\right).
 \end{array}    \right. 
 \end{eqnarray}
 \noindent  By using the present solution, the propagation speeds of slow waves are not dependent on the intermediate states within the Riemann fan. The present solution is of course not unique, and more rigorous solutions may be provided, which are, however, beyond the scope of this work.

An important note: when solving the conservation form of the MHD equations, numerical solvers may tolerate slightly negative thermal pressure \cite{zhang2026}, which does not necessarily sabotage the calculation of the fast magnetoacoustic speed. More specifically, when the plasma $\beta$ is small, in which case the positivity of scalar variables is usually prone to break, the fast speed is also relatively large such that a small negative pressure would not cause any real damage. However, the slow magnetoacoustic speed can be very small and even reach zero, so negative pressure can easily break the slow-wave calculation. 

Therefore, we currently use the HLLD-ec scheme \cite{zhang2026} to replace the HLLx scheme in locations where negative thermal pressure appears, and we discuss the resulting effects in the next section. Fully ensuring the positivity of scalar variables is not trivial \cite{Wu_2018,TREMBLIN2024,zhang2026}, and we do not discuss it further in this work. 

\section{Numerical tests} \label{sec:tests}

To evaluate the proposed scheme, we compare it mainly with the HLLD scheme and, in some cases, the Roe scheme \cite{Roe1981,BRIO1988,Balsara1998}. Note that in the implementation of the HLLx scheme, a small value may be used when a formula becomes indeterminate. We omit these details for simplicity.  We use an explicit FV solver\footnote{Opensource package  \texttt{MLAU} \cite{Minoshima_2019,Minoshima2020} available at {https://github.com/minoshim/MLAU}.}, in which the divergence constraint is ensured by a constrained transport method \cite{Evans_1988,Toth_2000}. Only first-order accurate spatial discretisation is used below to exclude potential oscillations due to higher-order approximations \cite{Zhang2024}. We integrate in time using the first-order forward Euler method.

\begin{figure}[h]
 \centering
 \subfigure[\label{fig:BW_rho}{}]{
 \includegraphics[width=0.48\textwidth]{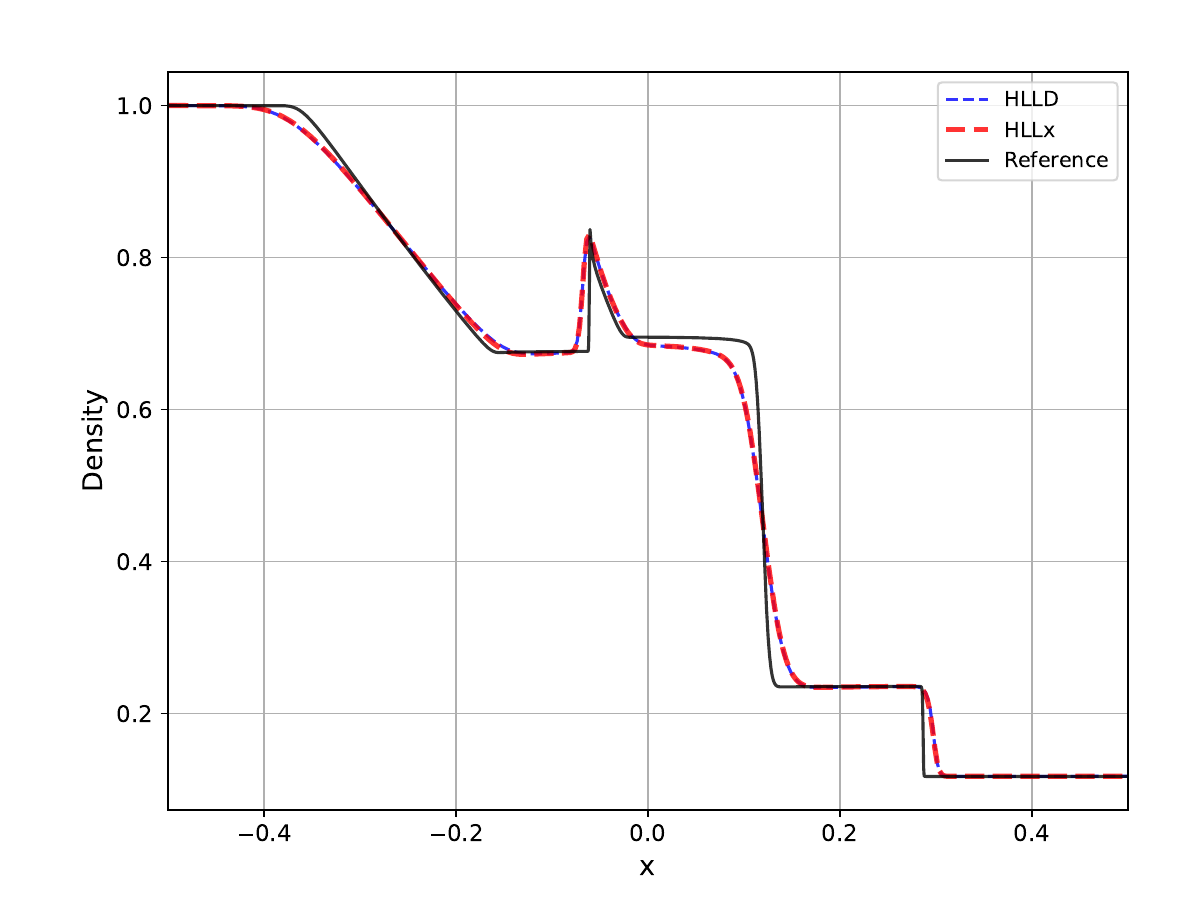}
 }
 \subfigure[\label{fig:BW_p}{}]{
 \includegraphics[width=0.48\textwidth]{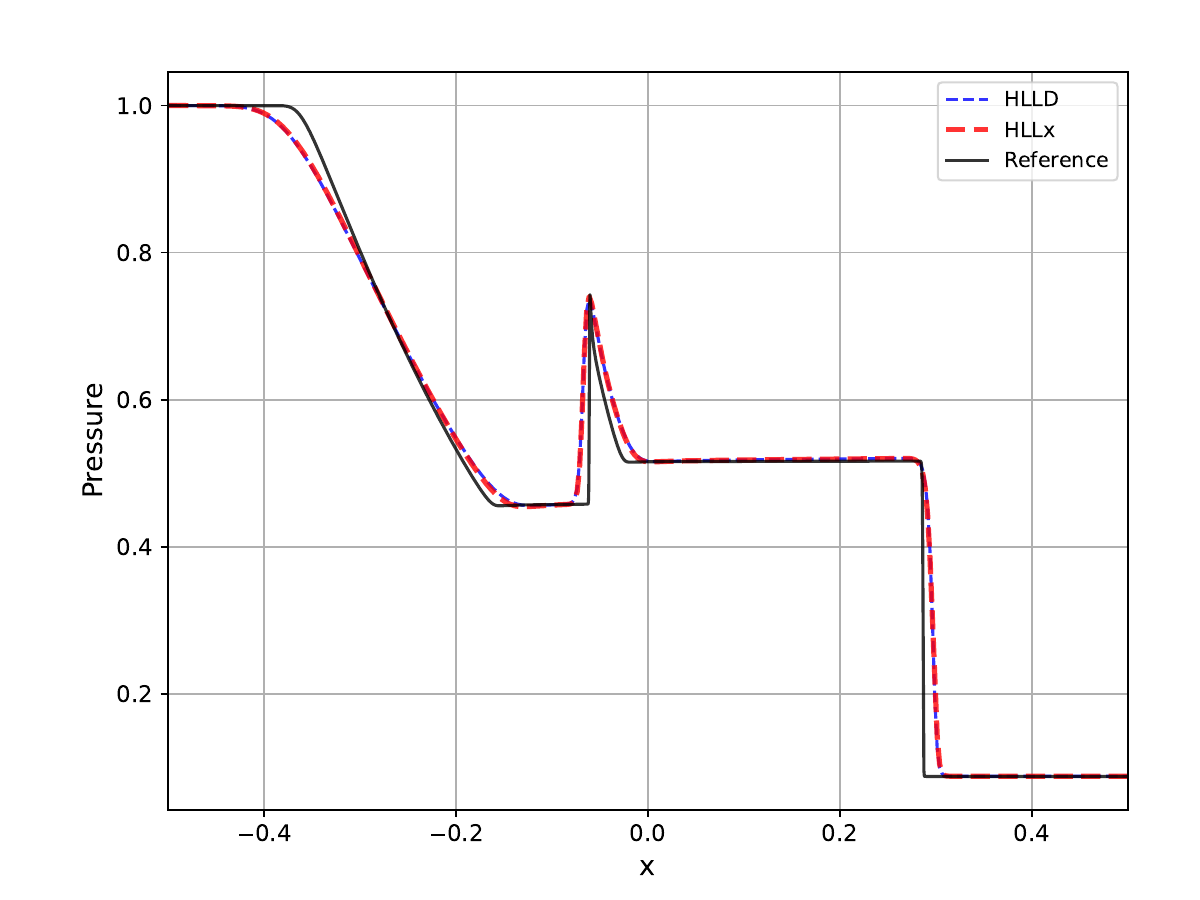}
 } 
\subfigure[\label{fig:BW_u}{}]{
 \includegraphics[width=0.48\textwidth]{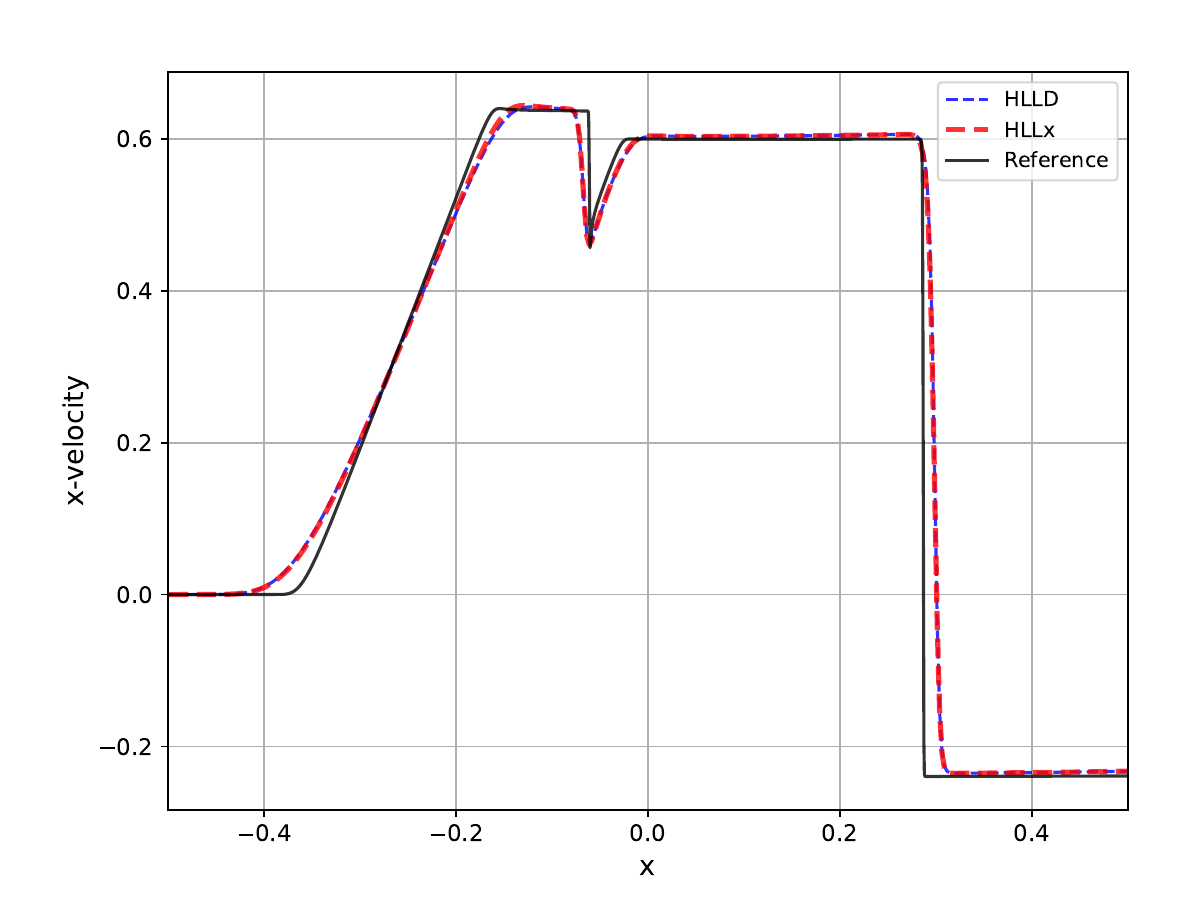}
 }
 \subfigure[\label{fig:BW_By}{}]{
 \includegraphics[width=0.48\textwidth]{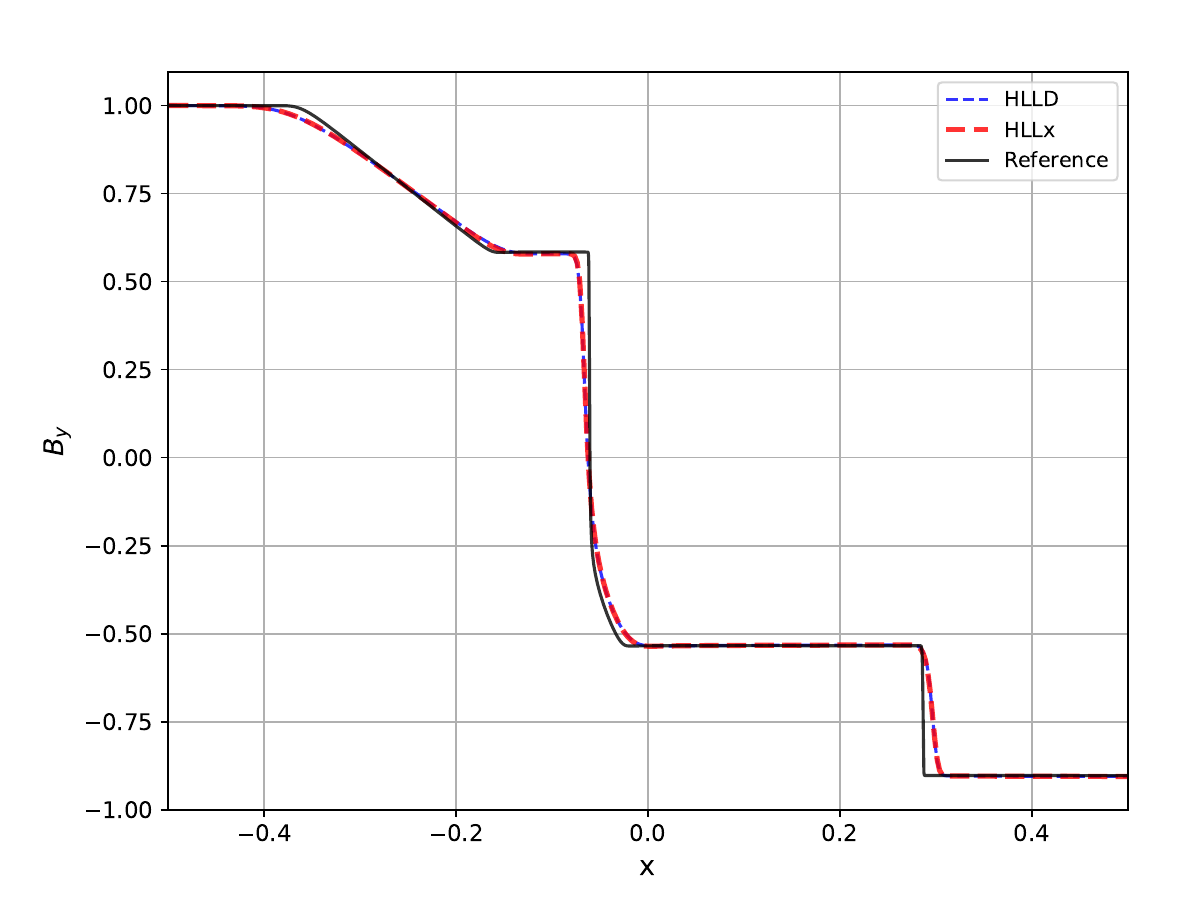}
 } 
 \caption{Results of the Brio-Wu shock tube problem.}
 \label{fig:BW}
\end{figure}

The HLLx scheme is improved mainly by including the slow mode; thus, when slow waves are important, the results of the HLLx scheme differ clearly from those of the HLLD scheme. In the following, we will emphasise this issue when discussing the numerical results.

\subsection{The 1D shock-tube problem of Brio and Wu} \label{sec:BW}

The MHD shock-tube problem of Brio and Wu \cite{BRIO1988} is first used to show the basic behaviors of the proposed scheme. This 1D problem within $x\in[-0.5,0.5]$ has two sets of initial states separated by a discontinuity, with a constant $B_x=0.75$ and an adiabatic index $\gamma=2$. The left and right states are:  
  \begin{eqnarray}   
  \left\{\begin{array}{c}
 (\rho, u, v, p, B_y, B_z)^{\text{l}}=(1, 0, 0, 1, 1, 0), \\
 (\rho, u, v, p, B_y, B_z)^{\text{r}}=(0.125, 0, 0, 0.1, -1, 0), 
 \end{array}    \right.
 \end{eqnarray} 
\noindent  A 1D mesh with $400$ grid cells is used to model this problem. A constant CFL$=0.4$ is used to reach non-dimensional time $t=0.2$. We provide a reference solution using the HLLx scheme with 4000 grid cells. The Brio-Wu problem creates fast rarefaction waves, a slow shock, a central contact discontinuity, and a slow compound wave.

The results are given in Figure \ref{fig:BW}. As the slow mode is not dominant in this case, the difference between the HLLD and HLLx schemes is minor. However, we still observe that the resolution of the HLLx scheme is slightly closer to the reference solution at the tail of the rarefaction wave. This is because rarefaction is not an eigenmode, and the HLLD scheme treats such variation as only the fast mode for compressible variations, leading to higher numerical diffusion than the HLLx scheme.

\begin{figure}[h]
 \centering
 \subfigure[\label{fig:DW_rho}{}]{
 \includegraphics[width=0.48\textwidth]{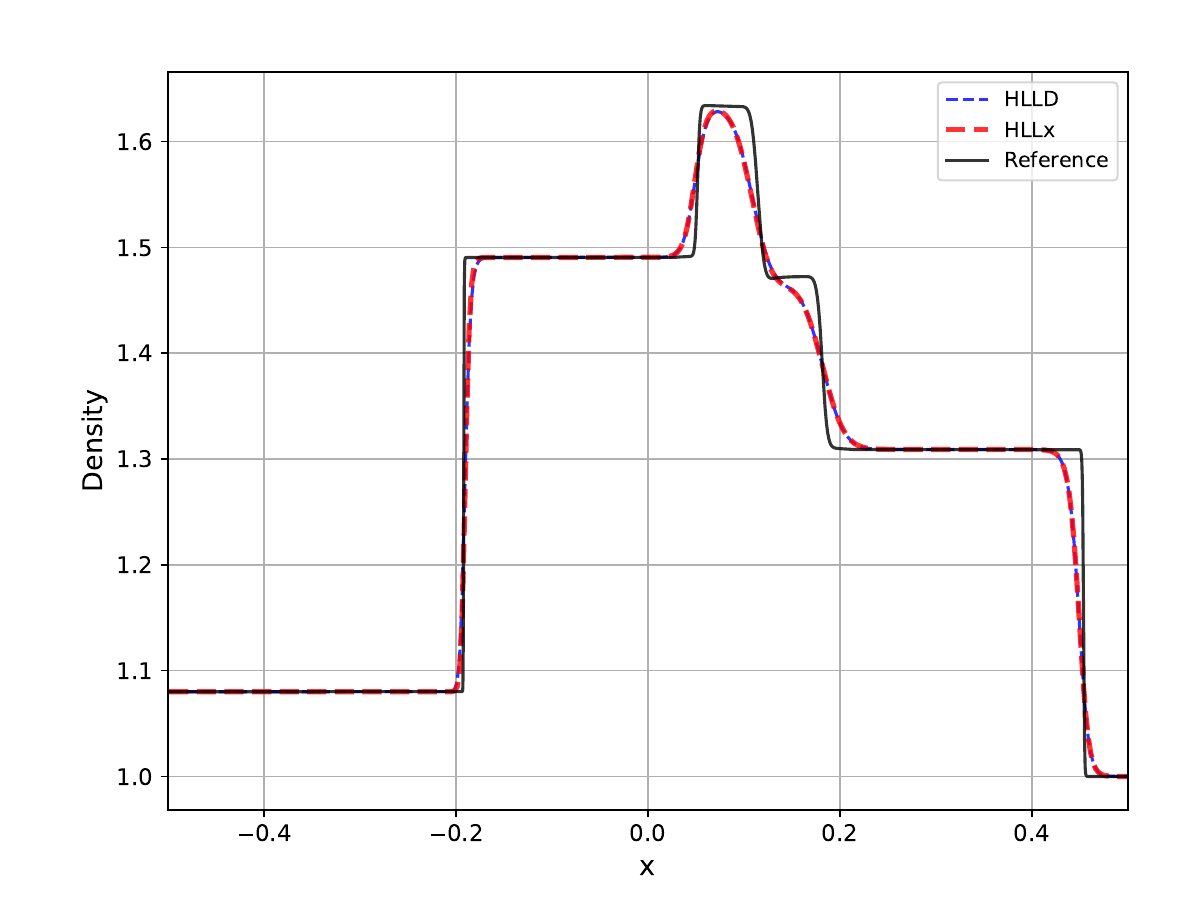}
 }
 \subfigure[\label{fig:DW_p}{}]{
 \includegraphics[width=0.48\textwidth]{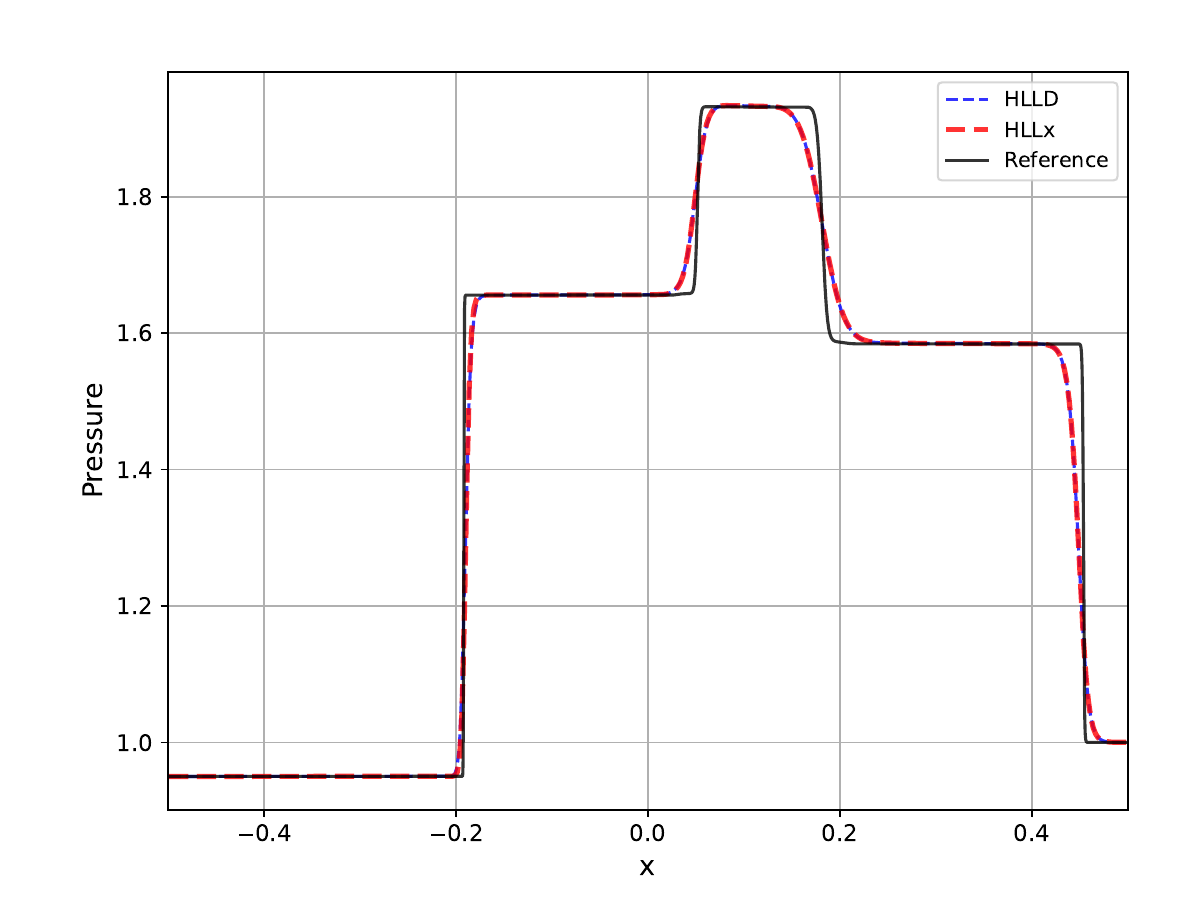}
 } 
\subfigure[\label{fig:DW_u}{}]{
 \includegraphics[width=0.48\textwidth]{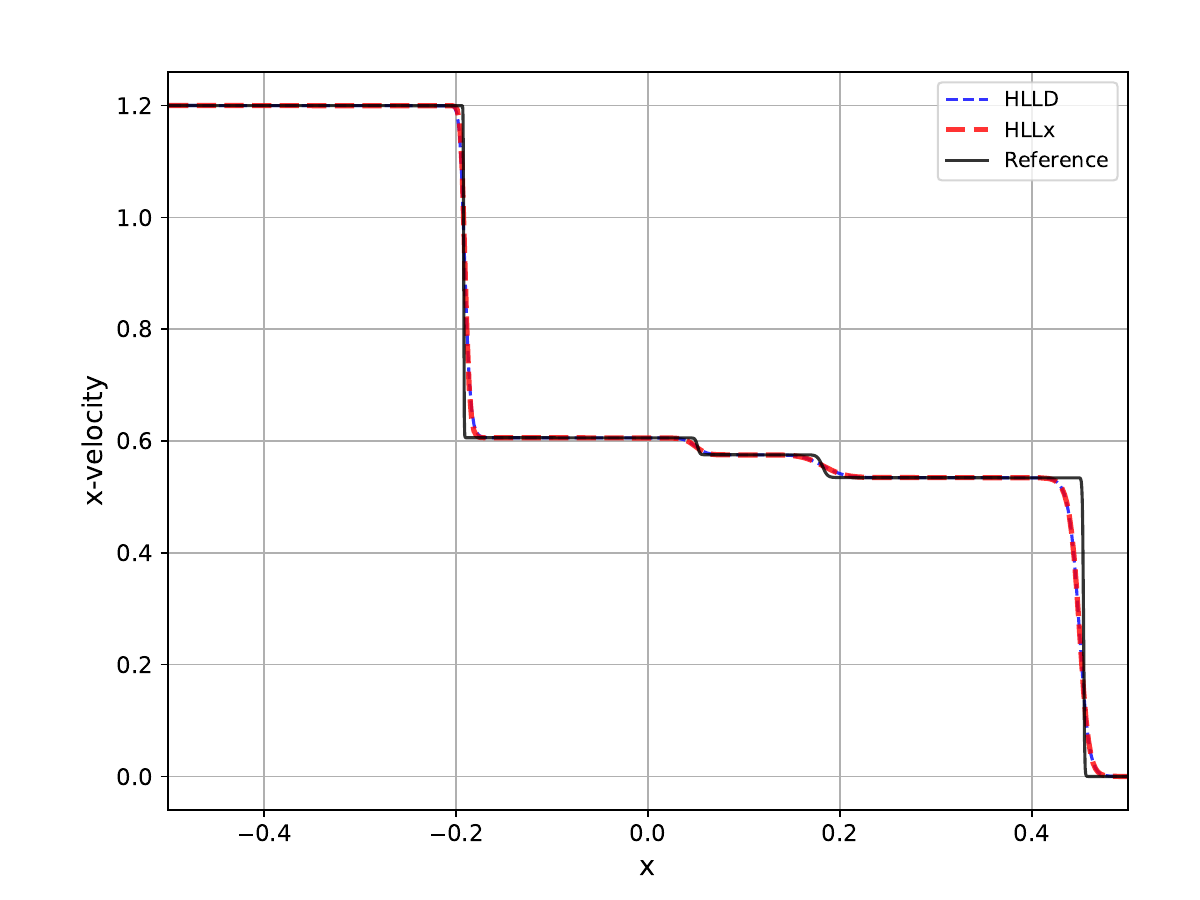}
 }
 \subfigure[\label{fig:DW_By}{}]{
 \includegraphics[width=0.48\textwidth]{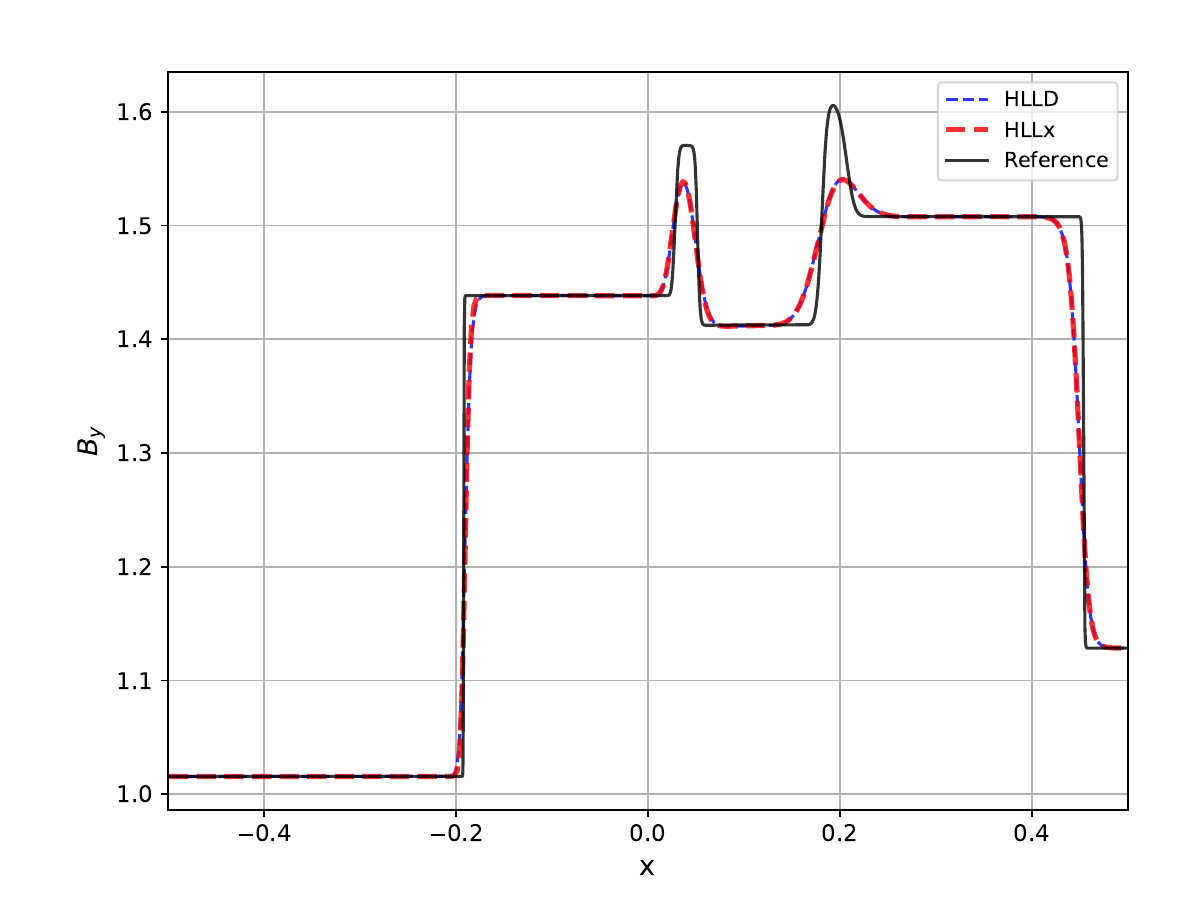}
 } 
 \caption{Results of the Dai-Woodward shock tube problem.}
 \label{fig:DW}
\end{figure}

\subsection{The 1D shock-tube problem of Dai and Woodward}
The MHD shock-tube problem originally proposed by Dai and Woodward \cite{DAI1994} and slightly revised in Ref.~\cite{Minoshima2020} is also tested, since this problem involves all 7 MHD waves in the MHD Riemann problem.  This 1D problem within $x\in[-0.5,0.5]$ has two sets of initial states separated by a discontinuity, with a constant $B_x=2/\sqrt{4\pi}$ and an adiabatic index $\gamma=5/3$. 
The left and right states are:  \begin{eqnarray}   
  \left\{\begin{array}{c}
 (\rho, u, v, w, p, B_y, B_z)^{\text{l}}=(1.08, 1.2, 0.01, 0.5, 0.95, 3.6/\sqrt{4\pi}, 2/\sqrt{4\pi}), \\
 (\rho, u, v, w, p, B_y, B_z)^{\text{r}}=(1, 0, 0, 0, 1, 4/\sqrt{4\pi}, 2/\sqrt{4\pi}). 
 \end{array}    \right.
 \end{eqnarray} 
 \noindent  Other details of the computational setting are the same as in the previous case.

The results at $t=0.2$ are shown in Figure \ref{fig:DW}. Again, the difference between the HLLD and HLLx schemes is minor. A minor difference can only be observed behind the left going fast shock, for which the HLLx scheme is slightly less diffusive. Therefore, when $B_x$ is weak, we may conclude that the HLLD and HLLx schemes perform similarly.

\begin{figure}[h]
 \centering
 \subfigure[\label{fig:StrongDW400_rho}{}]{
 \includegraphics[width=0.48\textwidth]{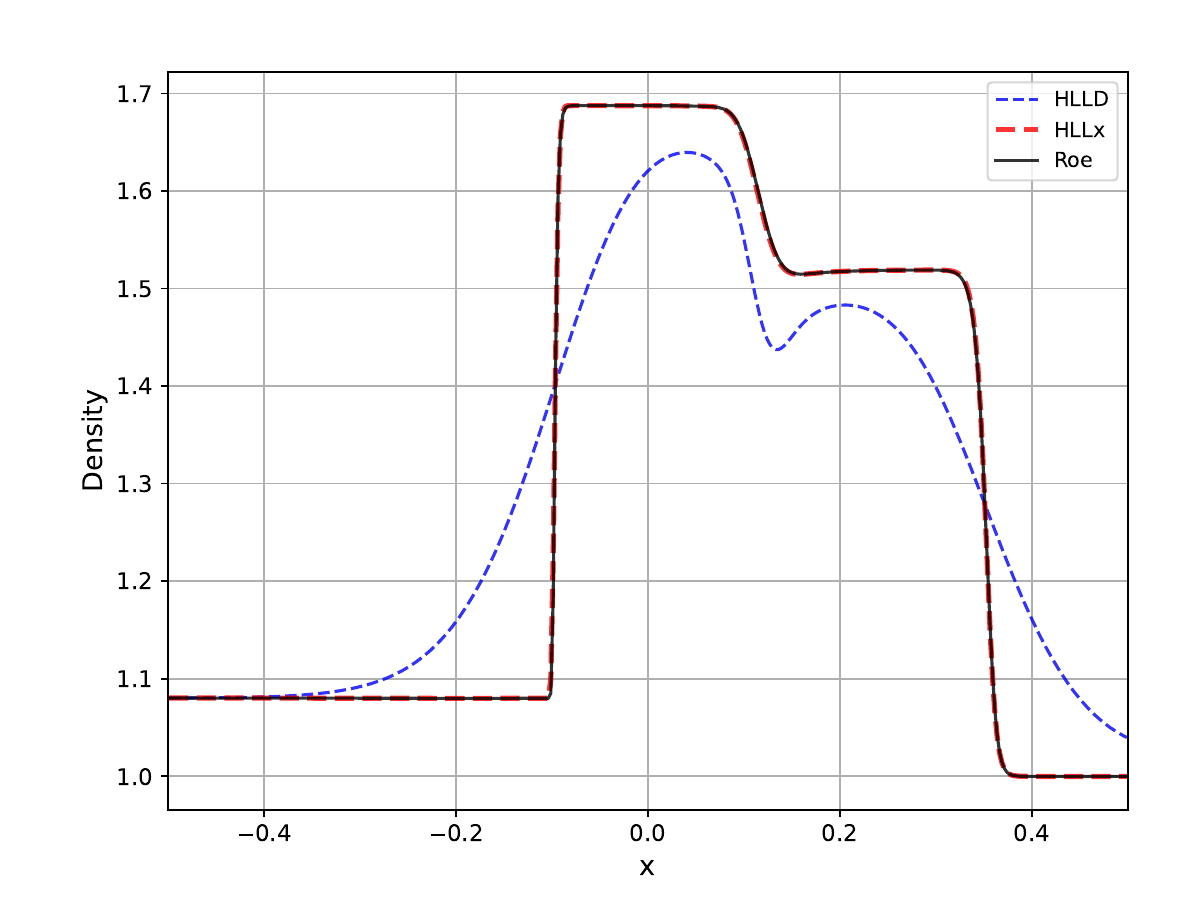}
 }
 \subfigure[\label{fig:StrongDW400_p}{}]{
 \includegraphics[width=0.48\textwidth]{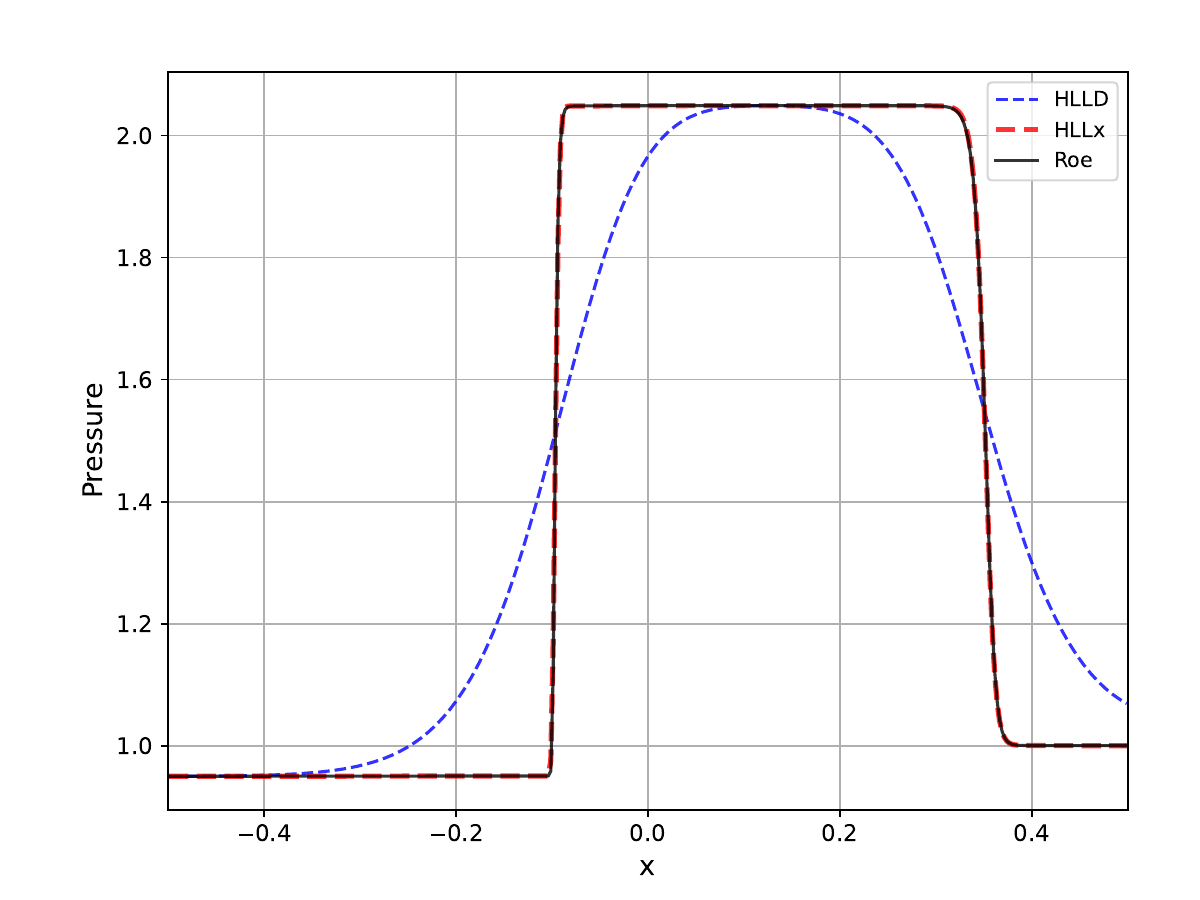}
 } 
\subfigure[\label{fig:StrongDW400_u}{}]{
 \includegraphics[width=0.48\textwidth]{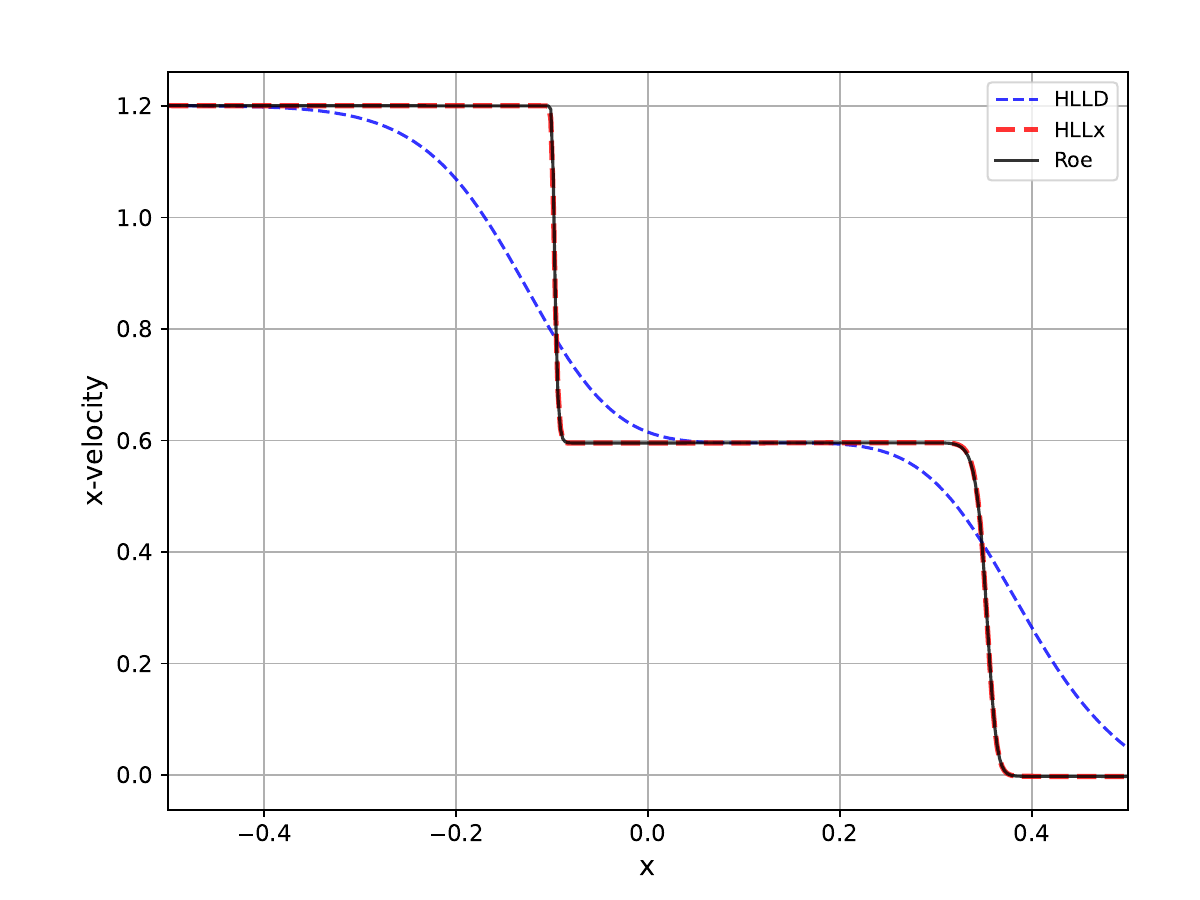}
 }
 \subfigure[\label{fig:StrongDW400_By}{}]{
 \includegraphics[width=0.48\textwidth]{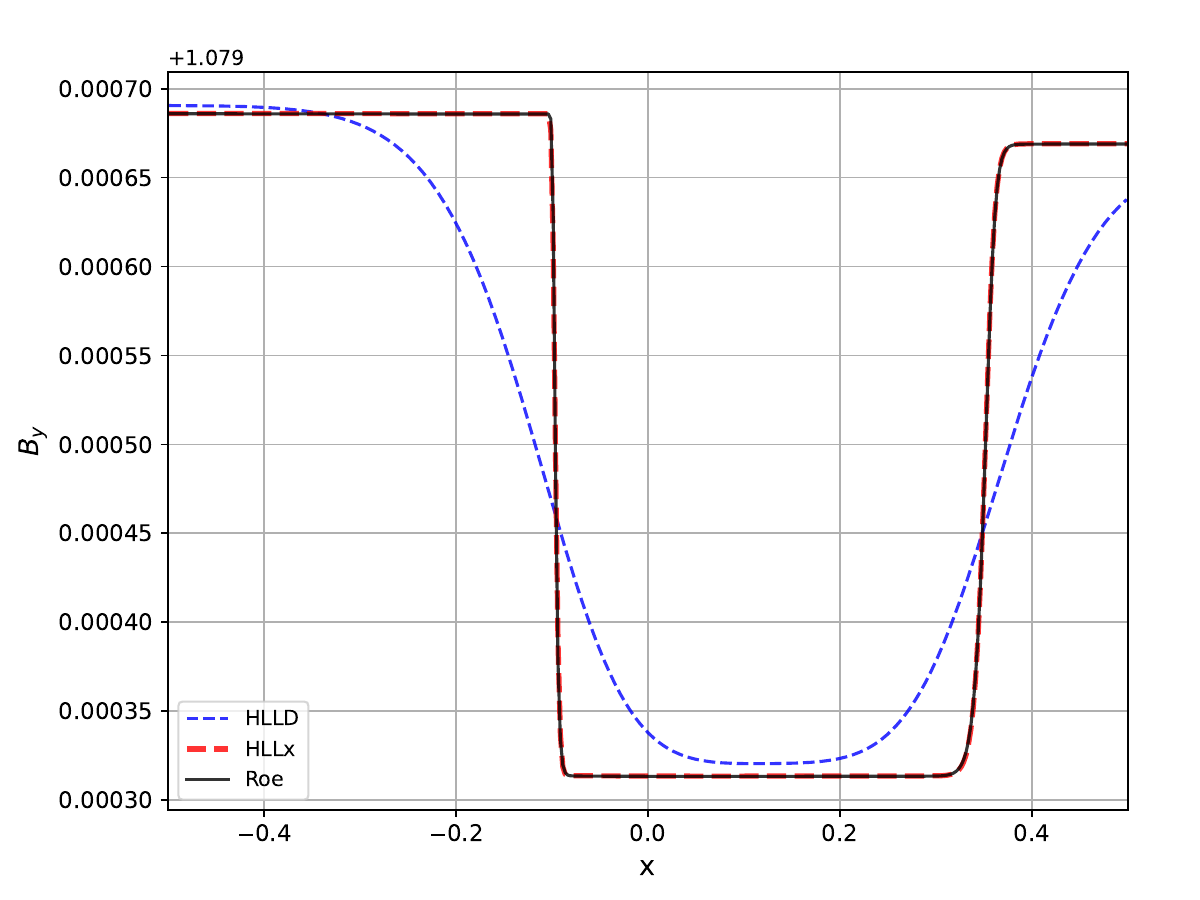}
 } 
 \caption{Results of the Dai-Woodward shock tube problem with large $B_x$: 400 grid cells.}
 \label{fig:StrongDW400}
\end{figure}

\subsection{The Dai and Woodward shock-tube problem with large $B_x$}

\begin{figure}[h]
 \centering
 \subfigure[\label{fig:StrongDW4000_rho}{}]{
 \includegraphics[width=0.48\textwidth]{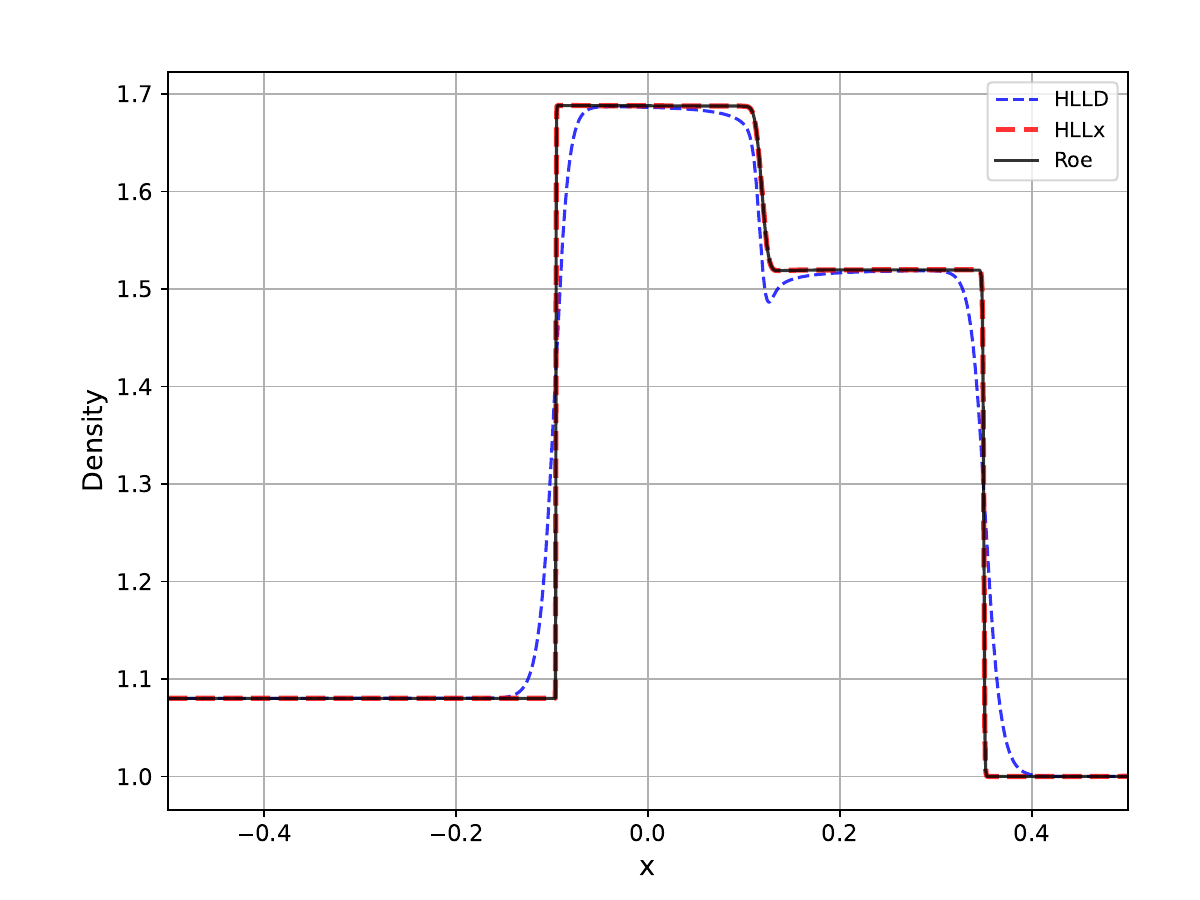}
 }
 \subfigure[\label{fig:StrongDW4000_p}{}]{
 \includegraphics[width=0.48\textwidth]{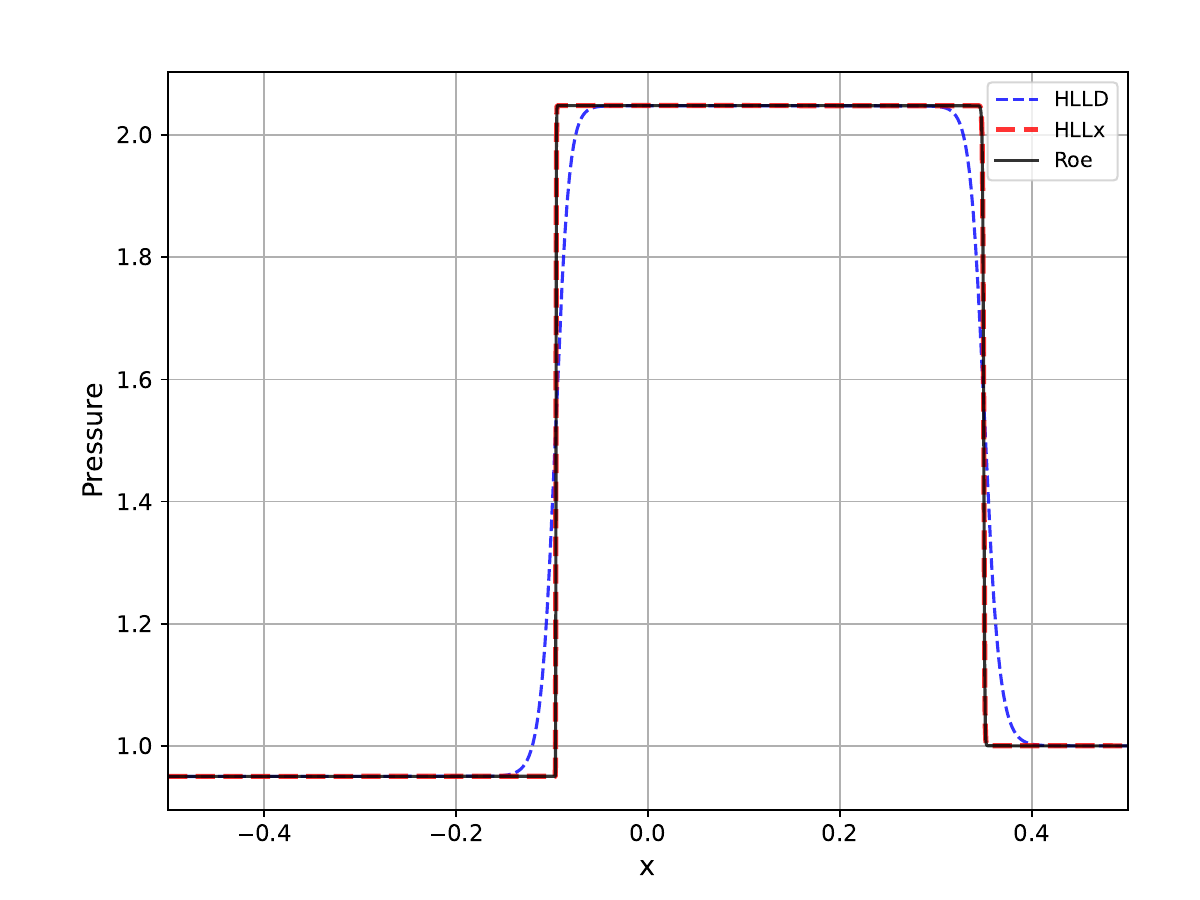}
 } 
\subfigure[\label{fig:StrongDW4000_u}{}]{
 \includegraphics[width=0.48\textwidth]{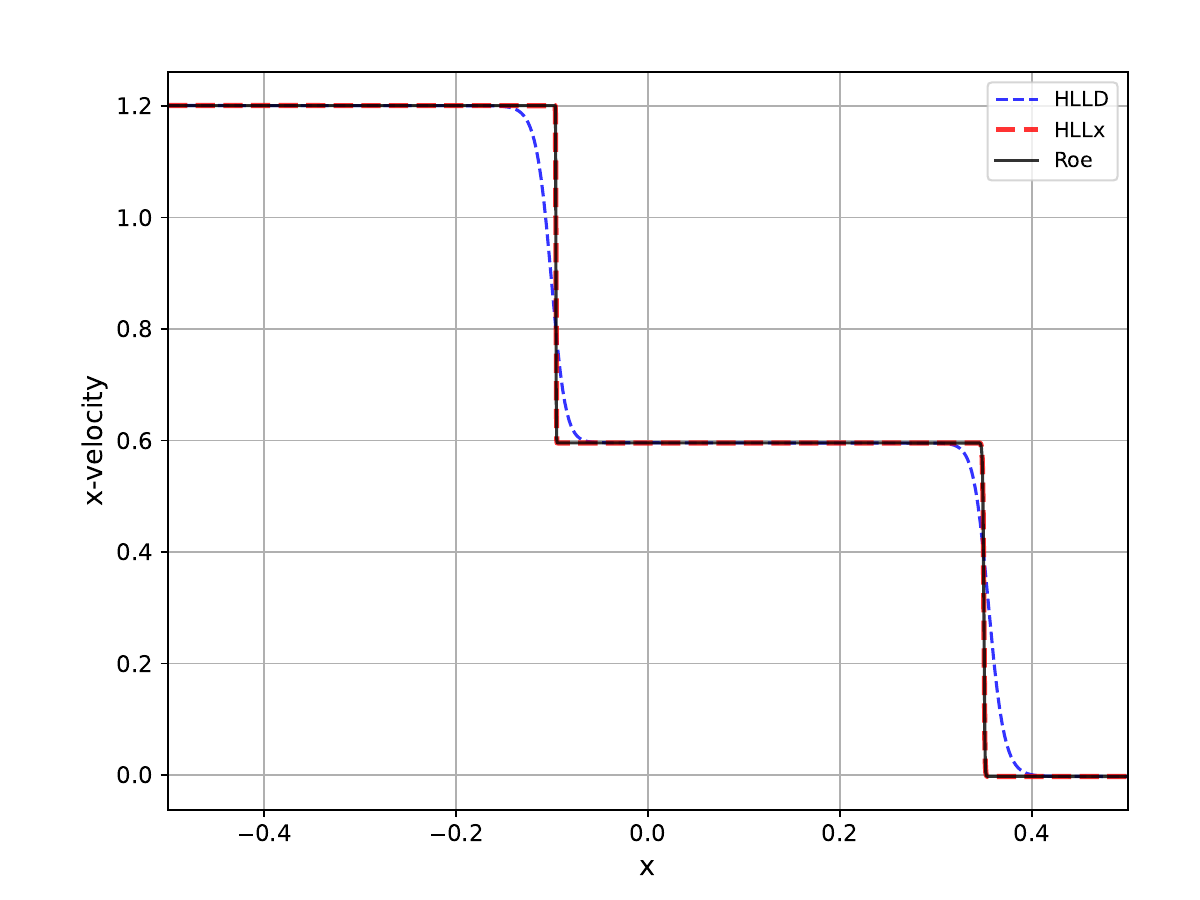}
 }
 \subfigure[\label{fig:StrongDW4000_By}{}]{
 \includegraphics[width=0.48\textwidth]{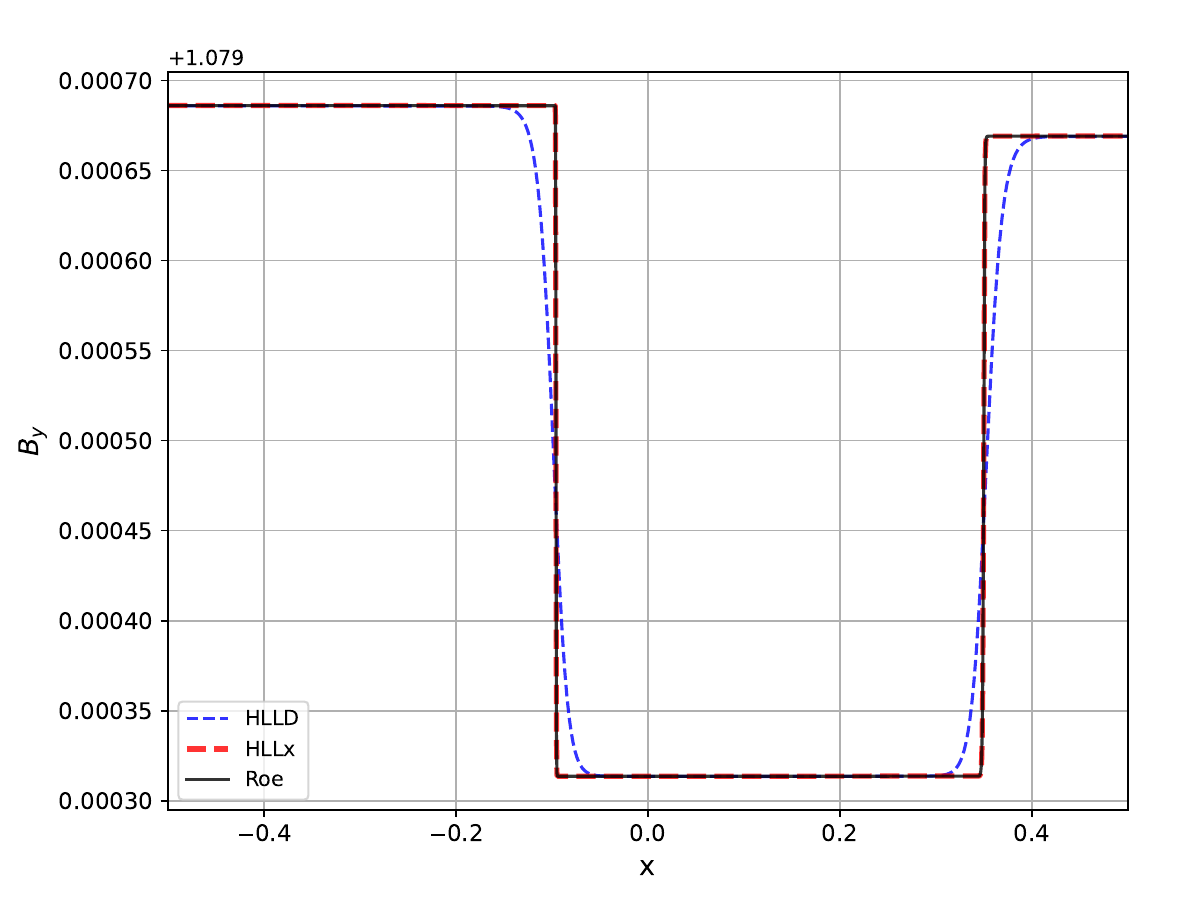}
 } 
 \caption{Results of the Dai-Woodward shock tube problem with large $B_x$: 4000 grid cells.}
 \label{fig:StrongDW4000}
\end{figure}

Originally presented in Ref. \cite{Minoshima2020}, we revise the shock tube problem based on the previous test case by changing the longitudinal magnetic field to $B_x=200/\sqrt{4\pi}$, without changing the other initial states. Therefore, we reuse the numerical setting. Additionally, we also consider the Roe scheme. Numerical results at $t=0.2$ on 400 grid cells are shown in Figure \ref{fig:StrongDW400}, and the results on 4000 grid cells are shown in Figure \ref{fig:StrongDW4000}. In this case, the fast and Alfv\'en modes are practically overlapped and travel away from the computational domain well before $t=0.2$.

In both figures, we can hardly observe any difference between the HLLx scheme and the Roe scheme.  In contrast, the HLLD scheme is clearly more diffusive. In fact, even on 4000 grid cells, the HLLD scheme still cannot capture the slow shocks as sharply as the HLLx and Roe schemes on 400 grid cells (except for the contact discontinuity). This is, however, expected, since we can calculate the eigenwave speeds and find that the fast magnetoacoustic speed is more than one order of magnitude faster than the slow magnetoacoustic speed, directly contributing to the numerical diffusion.

Providing a solution close(r) to that of the Roe scheme in this test case was not trivial \cite{Minoshima2020}, but the present HLLx scheme is able to produce almost identical resolution. Then a natural question is the difference between the HLLx scheme and the Roe scheme. We address this issue with the next test case.

\subsection{The slow switch-off rarefaction problem}

\begin{figure}[h]
 \centering
 \subfigure[\label{fig:SSR400_rho}{}]{
 \includegraphics[width=0.48\textwidth]{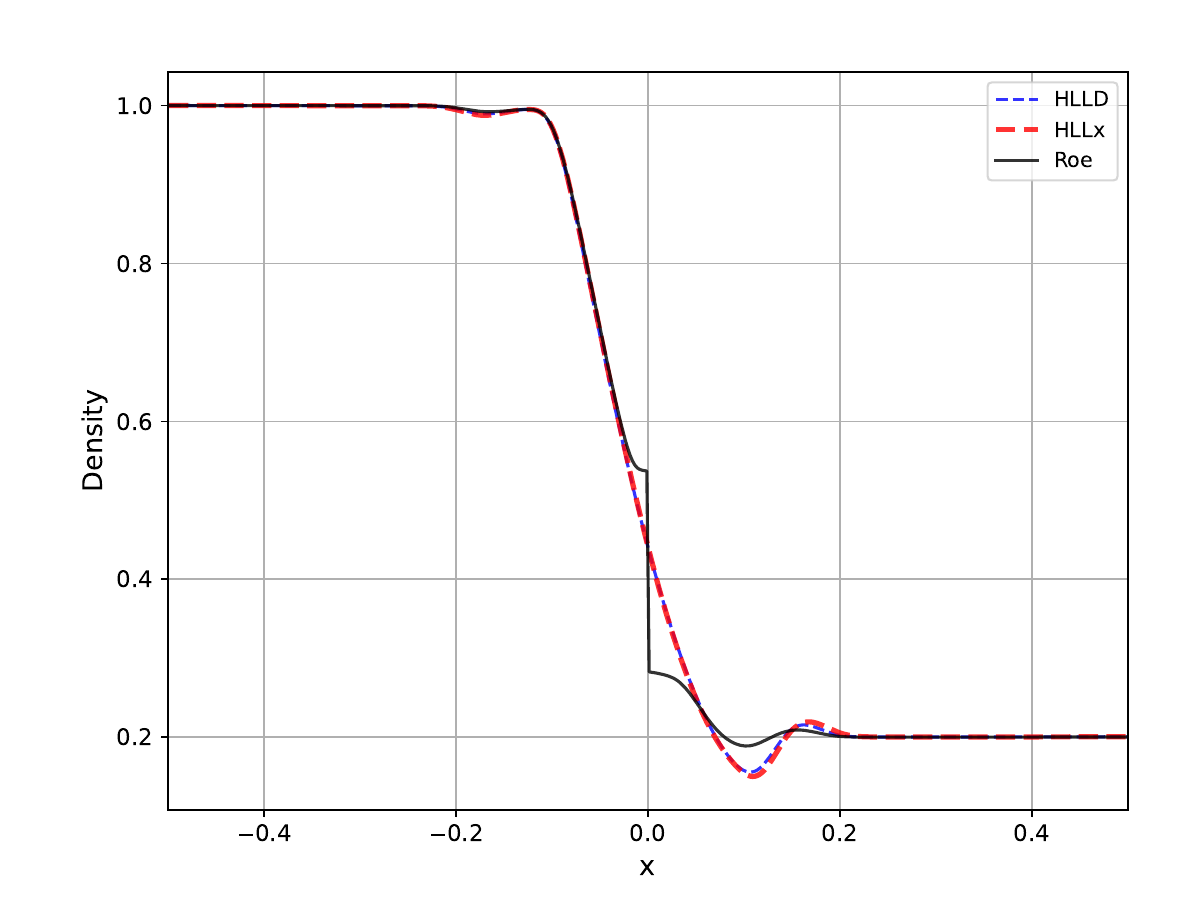}
 }
 \subfigure[\label{fig:SSR400_p}{}]{
 \includegraphics[width=0.48\textwidth]{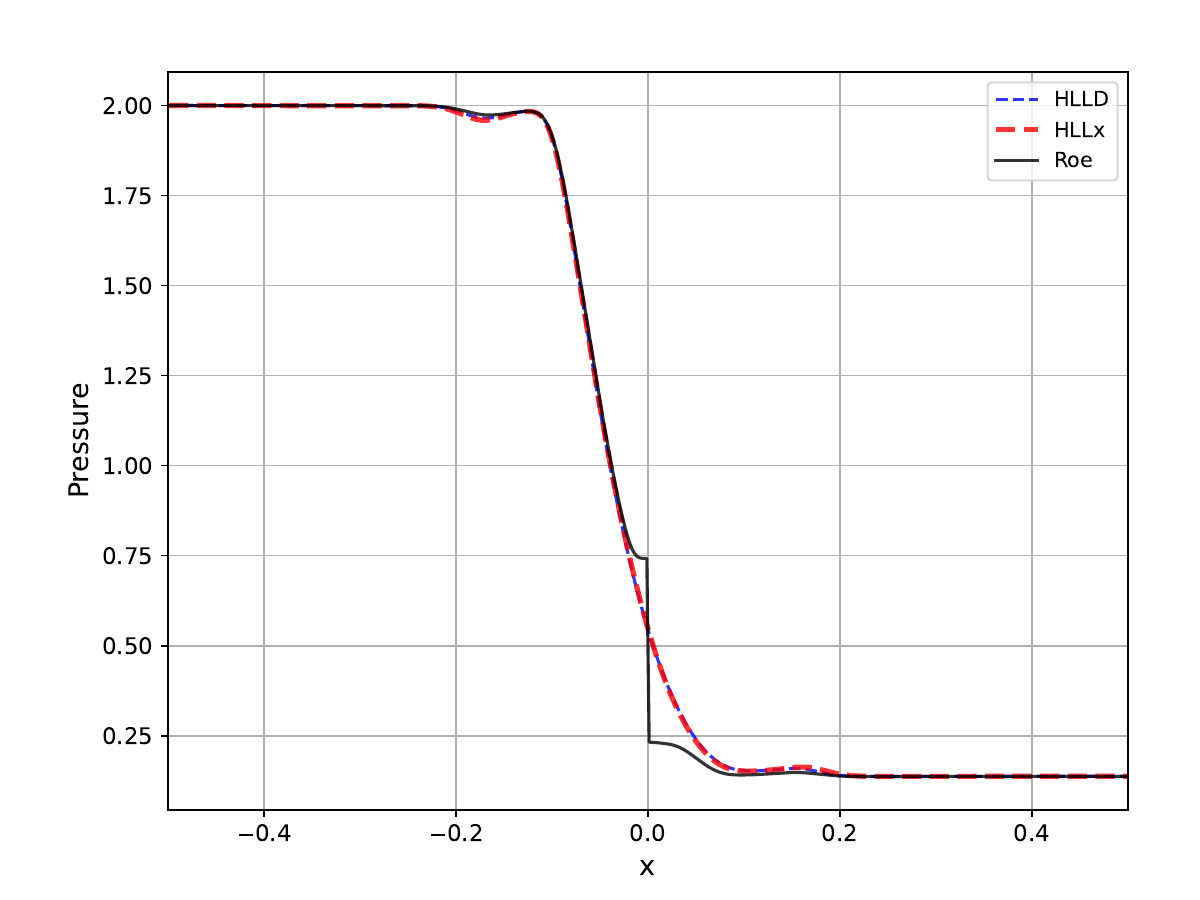}
 } 
\subfigure[\label{fig:SSR400_u}{}]{
 \includegraphics[width=0.48\textwidth]{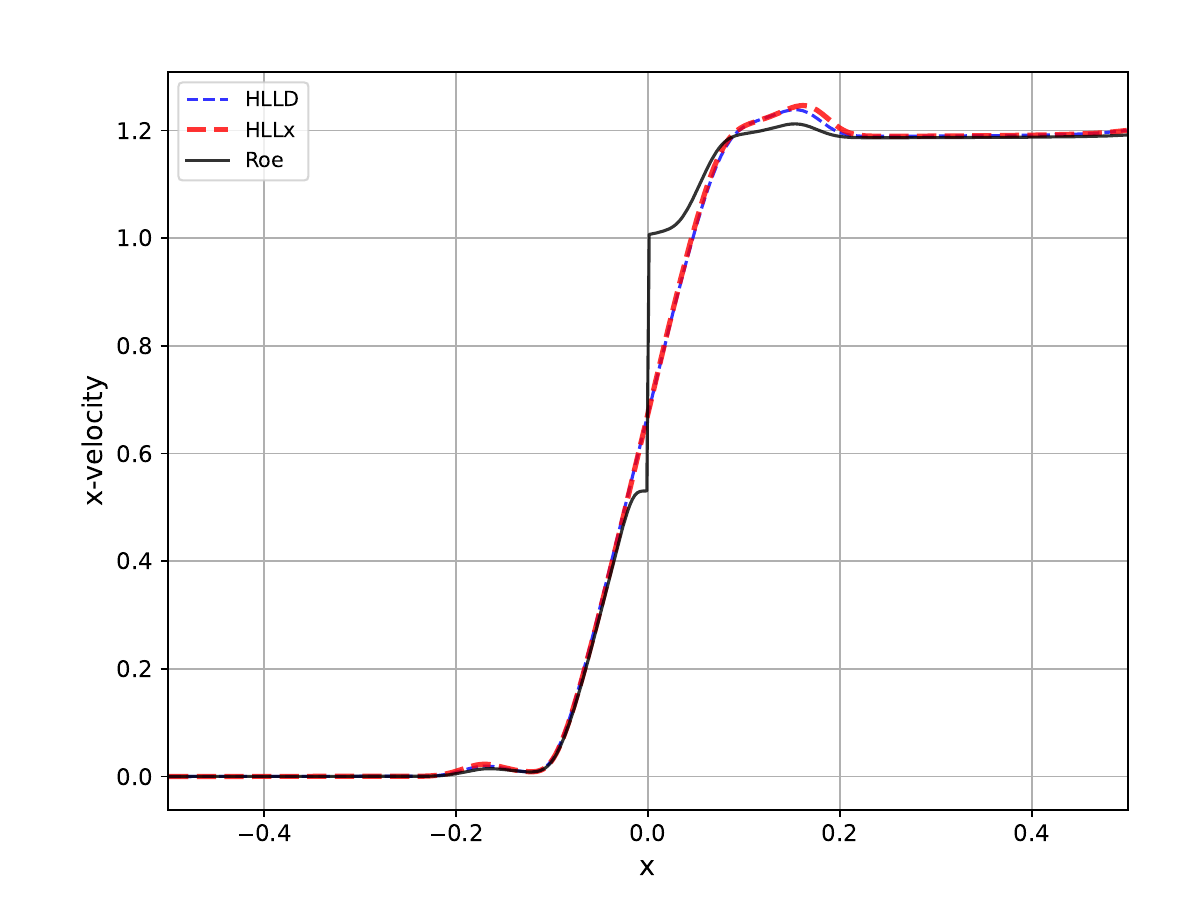}
 }
 \subfigure[\label{fig:SSR400_By}{}]{
 \includegraphics[width=0.48\textwidth]{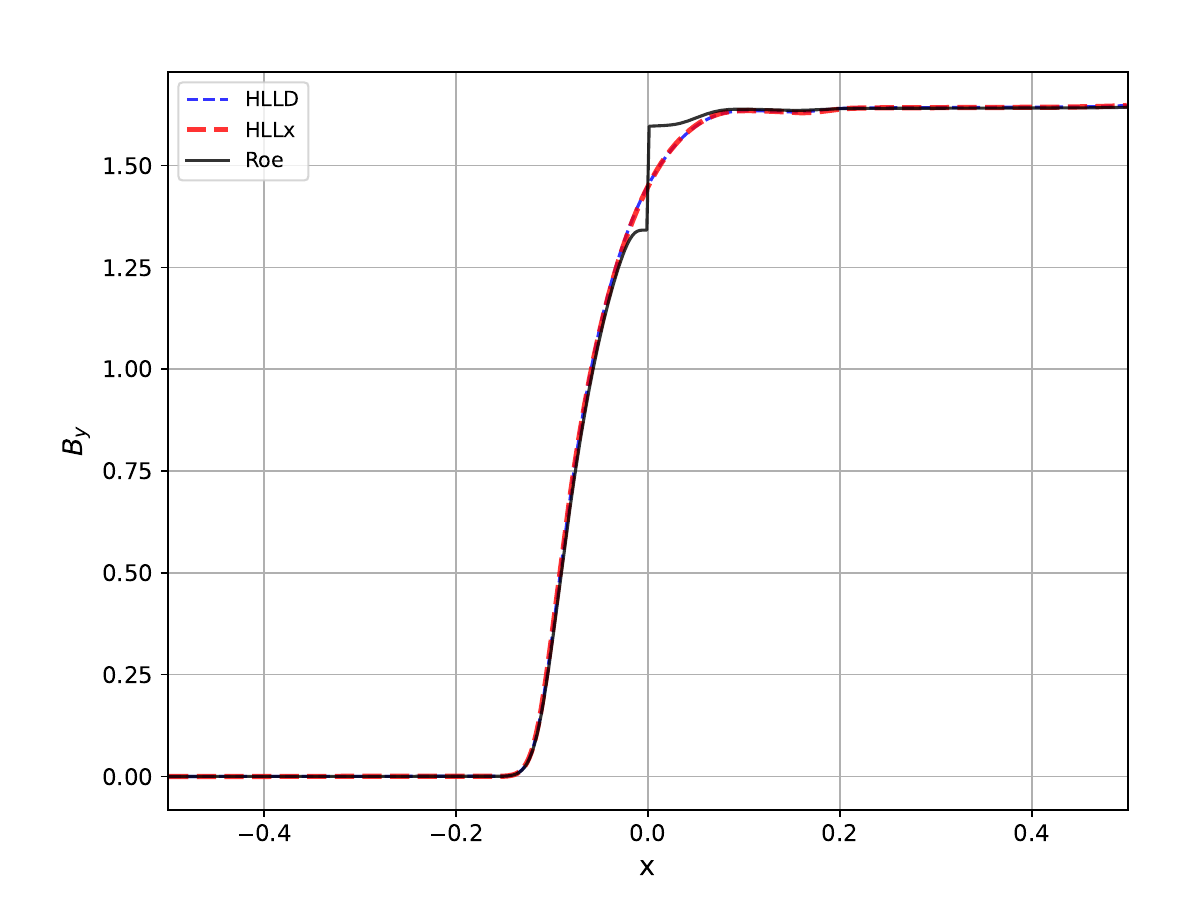}
 } 
 \caption{Results of the slow switch-off rarefaction problem.}
 \label{fig:SSR400}
\end{figure}

The challenge of the Roe scheme is its violation of the entropy condition. Therefore, here we use the slow switch-off rarefaction problem \cite{Miyoshi2005} to examine the performance of the numerical schemes. The two sets of initial states used are
\begin{eqnarray}   
  \left\{\begin{array}{c}
 (\rho, u, v, w, p, B_y, B_z)^{\text{l}}=(1, 0, 0, 0, 2, 0, 0), \\
 (\rho, u, v, w, p, B_y, B_z)^{\text{r}}=(1.2, 1.186, 2.967, 0, 0.1368, 0.6405, 0),
 \end{array}    \right.
 \end{eqnarray} 
\noindent with a constant $B_x=1$ and an adiabatic index $\gamma=5/3$. This initial condition will lead to a strong expansion, which is known to be challenging for the Roe scheme.
Numerical results at $t=0.1$ and on 400 grid cells are shown in Figure \ref{fig:SSR400}.

With a relatively small value of $B_x$, we again see only small differences between the HLLD and HLLx schemes. In comparison, the Roe scheme produces an expansion shock, which is clearly not physical.  An entropy fix would remove the expansion shock \cite{Miyoshi2005}, and this issue has been extensively discussed (e.g., Ref. \cite{HARTEN1983_2}). Nonetheless, here we may conclude that the present HLLx can be as accurate as the Roe scheme without risking a violation of the entropy condition.  

\subsection{The Orszag–Tang vortex problem}

\begin{figure}
 \centering
 \subfigure[\label{fig:t5_density_HLLD}{Density (HLLD)}]{
 \includegraphics[width=0.48\textwidth]{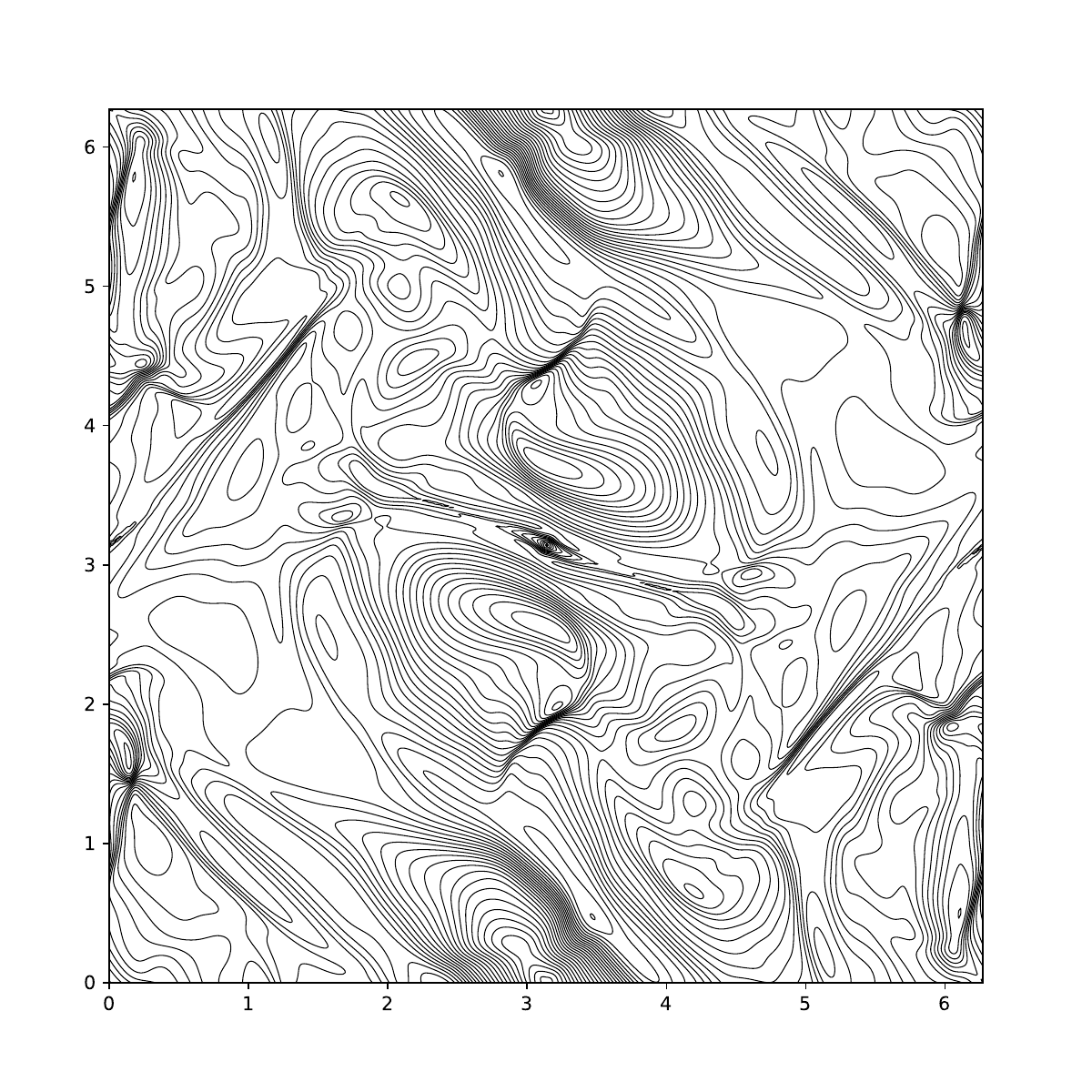}
 }
 \subfigure[\label{fig:t5_density_HLLX}{Density (HLLx)}]{
 \includegraphics[width=0.48\textwidth]{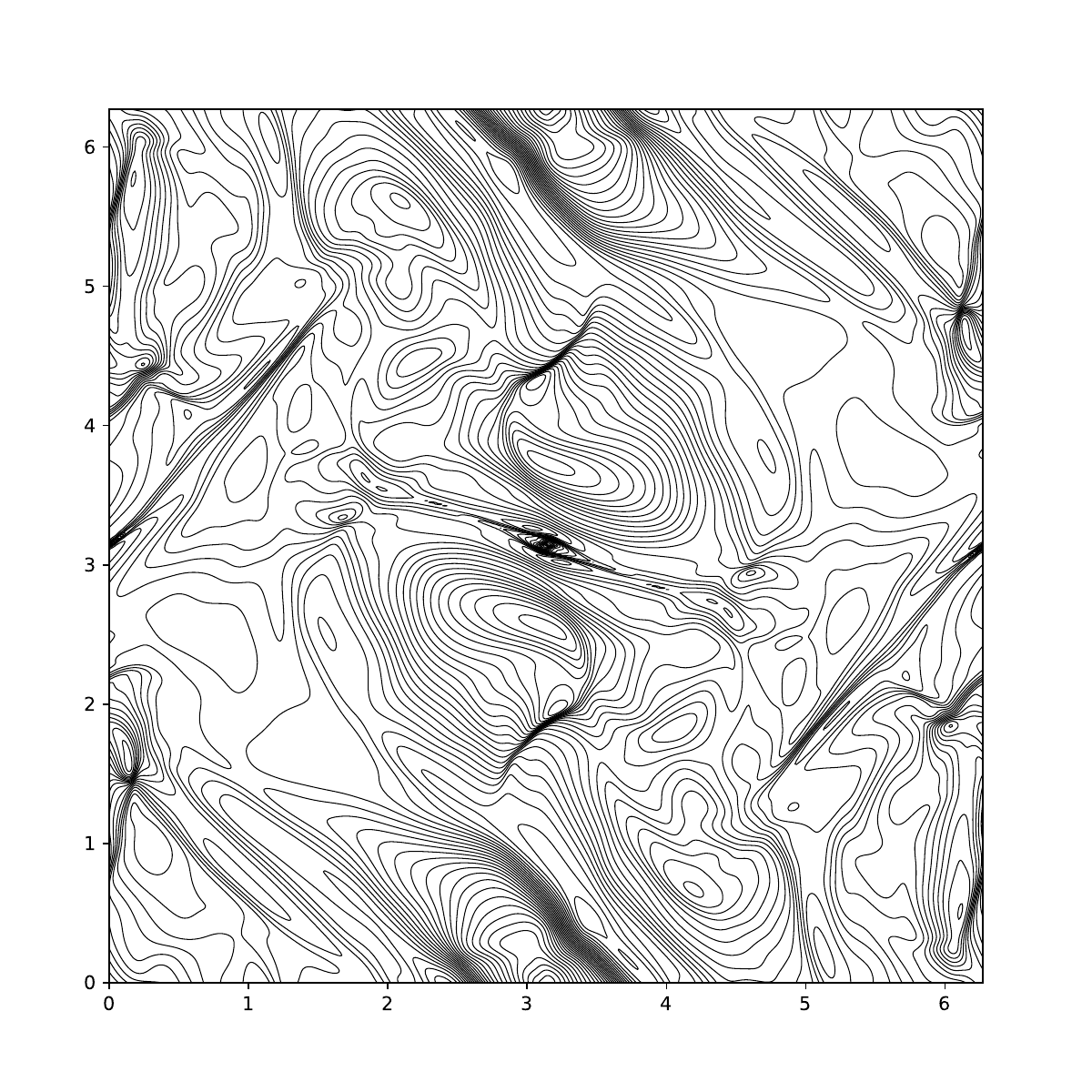}
 } 
\subfigure[\label{fig:t5_total_pressure_HLLD}{Total pressure (HLLD)}]{
 \includegraphics[width=0.48\textwidth]{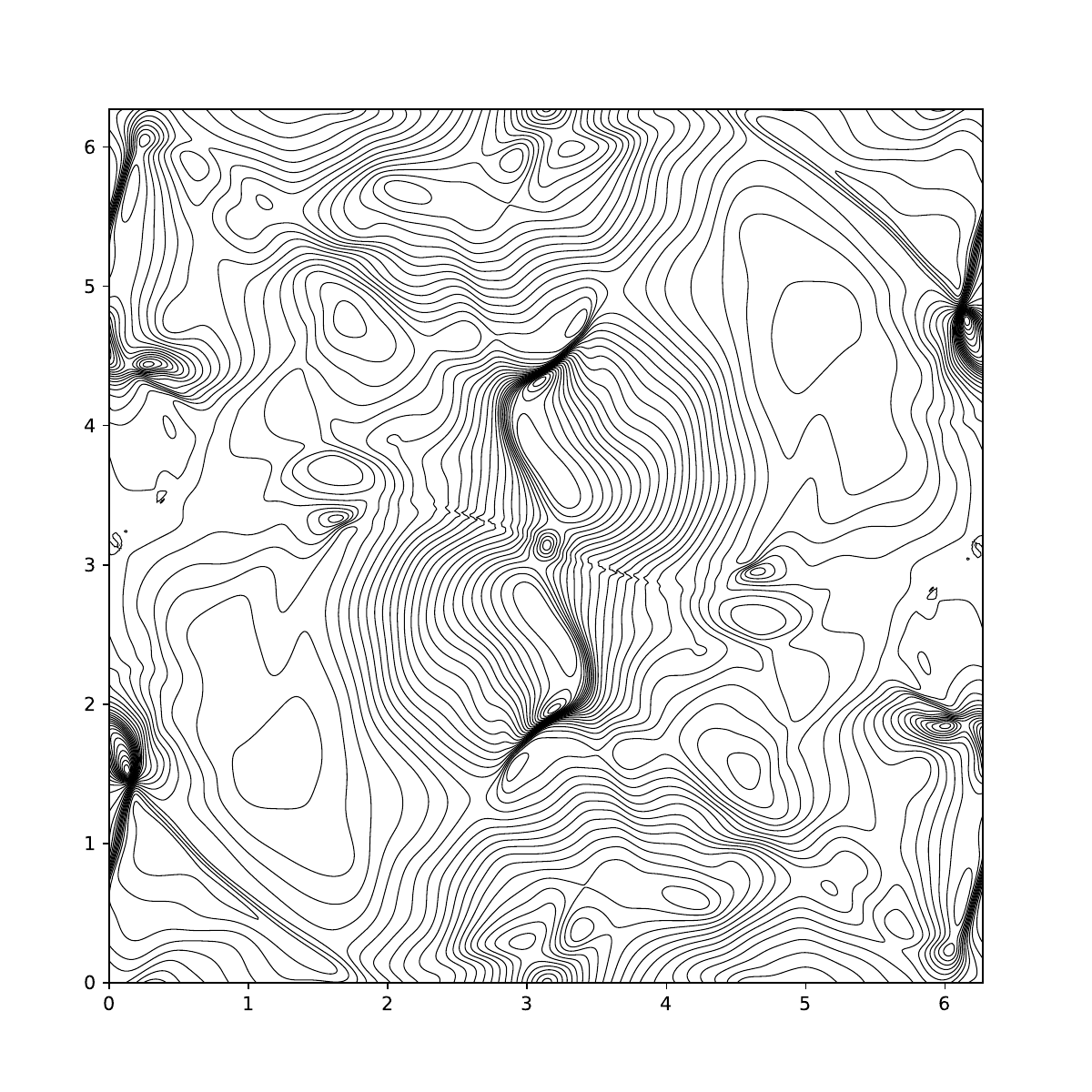}
 }
 \subfigure[\label{fig:t5_total_pressure_HLLX}{Total pressure (HLLx)}]{
 \includegraphics[width=0.48\textwidth]{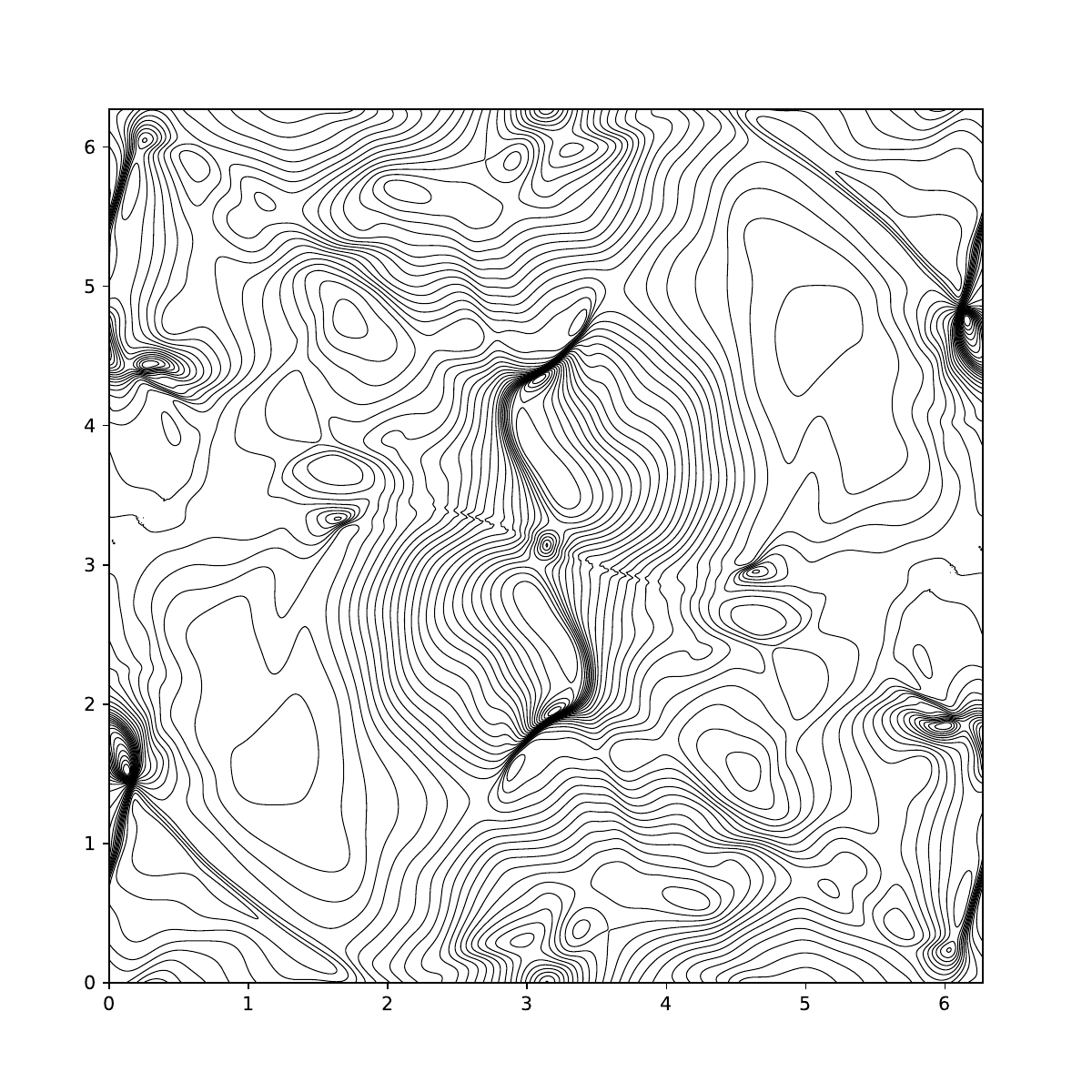}
 } 
 \caption{Results of the Orszag–Tang vortex problem.  In each plot 30 contour lines are shown, for density $\rho\in[1.13, 5.69]$, and for total pressure $P\in[1.72, 6.05]$, approximately.}
 \label{fig:OT_t5}
\end{figure}

\begin{figure}[h]
 \centering
 \subfigure[\label{fig:OT_density_slices_vertical}{}]{
 \includegraphics[width=0.48\textwidth]{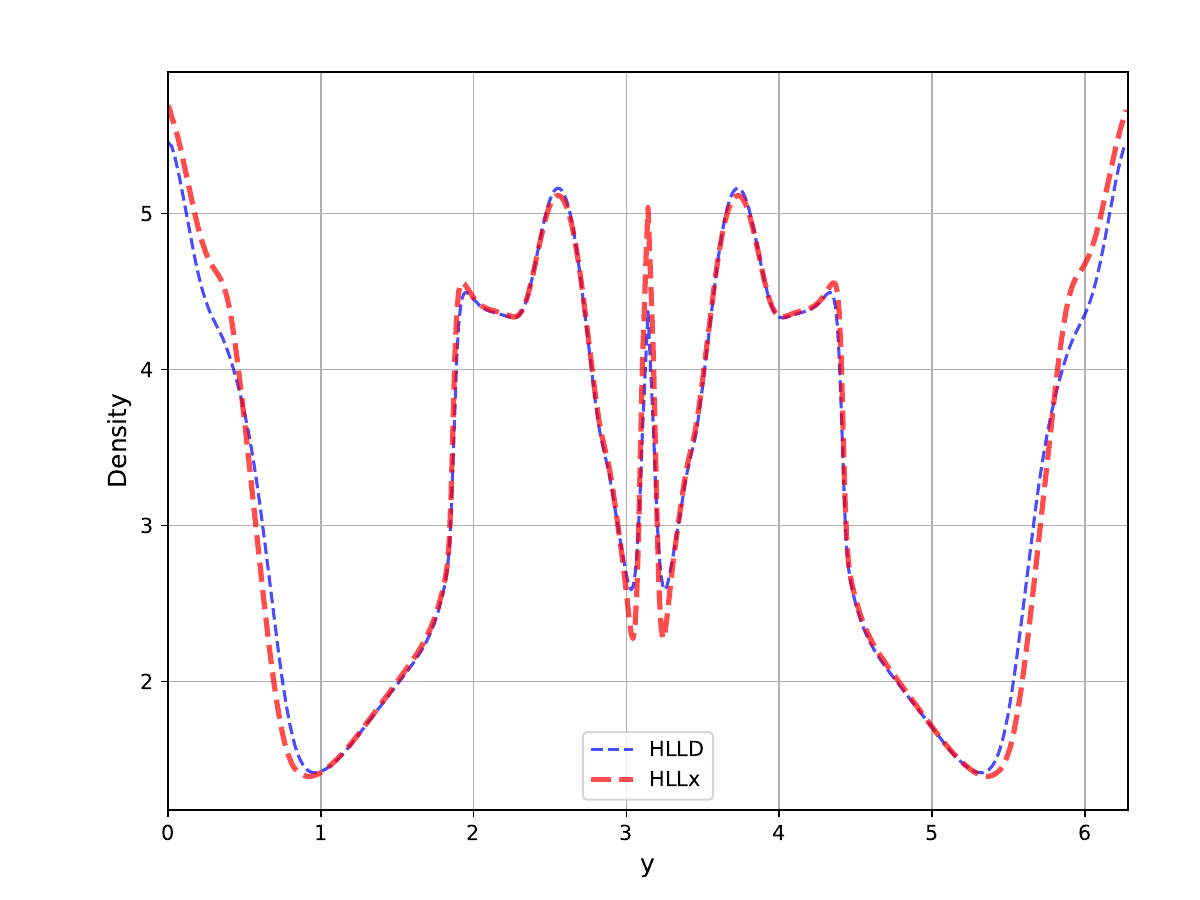}
 }
 \subfigure[\label{fig:OT_total_pressure_slices_vertical}{}]{
 \includegraphics[width=0.48\textwidth]{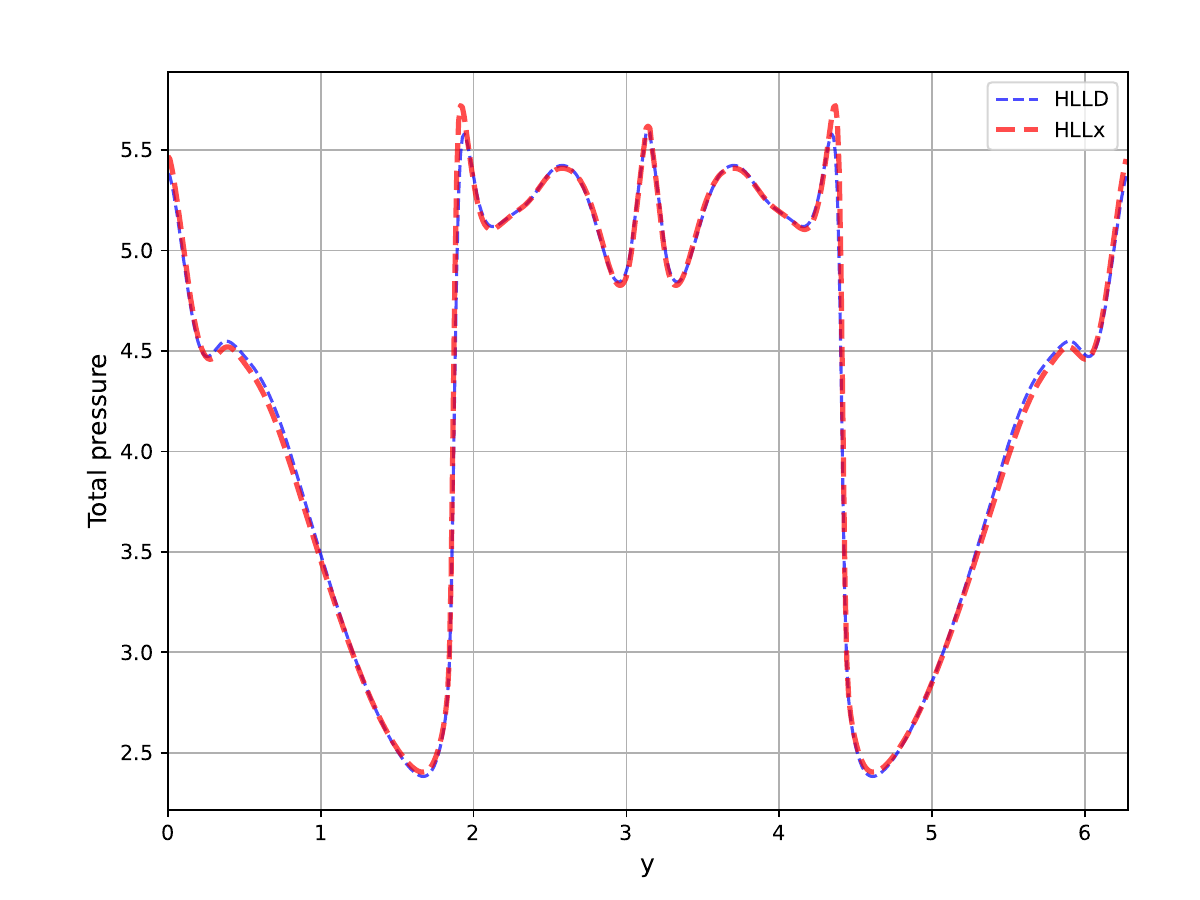}
 }  
 \caption{A vertical slice (at $x=\pi$) of the Orszag–Tang vortex problem.}
 \label{fig:OT_slices}
\end{figure}
 
The Orszag–Tang vortex problem \cite{Orszag1979} is frequently used to test MHD numerical methods because it develops interactions between MHD shock waves during the vortex evolution.  In this problem, although fast modes are dominant initially, slow modes soon develop \cite{Snow2021OT}, and thus we may observe the differences between the aforementioned numerical schemes.  
Here, the initial conditions are 
\begin{equation}
(\rho, u, v, w, p, B_x, B_y, B_z)=(\gamma^2, -\sin(y), \sin(x), 0, \gamma, -\sin(y), \sin(2x), 0),
\end{equation}
\noindent where $\gamma=5/3$, on a $[0,2\pi]\times[0,2\pi]$ domain, which is discretized by $512\times 512$ grid cells. We again use first-order accurate spatial and temporal schemes, with periodic boundary conditions and CFL$=0.4$.  Numerical results at $t=5$ are shown in Figures \ref{fig:OT_t5} and \ref{fig:OT_slices}.

This problem has a relatively high plasma $\beta$, and initially the fast mode is dominant. Therefore, the difference between the HLLD and HLLx schemes should not be significant. However, we still observe sharper small structures in the HLLx result as intermediate and slow shocks develop \cite{Snow2021OT}. The differences are more obvious in the density distributions and are still observable in the total pressure distributions. Interestingly, when separately visualising the distributions of thermal pressure and magnetic pressure, the results of the HLLx scheme are, in fact, smoother. They are, however, not further discussed since total pressure is a more relevant variable for MHD problems.

Here, we do not intend to discuss the detailed physics of the Orszag–Tang vortex problem, but we may claim that the improvement provided by the HLLx scheme could potentially be important for capturing small-scale structures in MHD turbulence simulations, even when plasma $\beta$ is relatively high.  

\subsection{The MHD blast problem}

\begin{figure}
 \centering
 \subfigure[\label{fig:t02_density_HLLD}{Density (HLLD)}]{
 \includegraphics[width=0.48\textwidth]{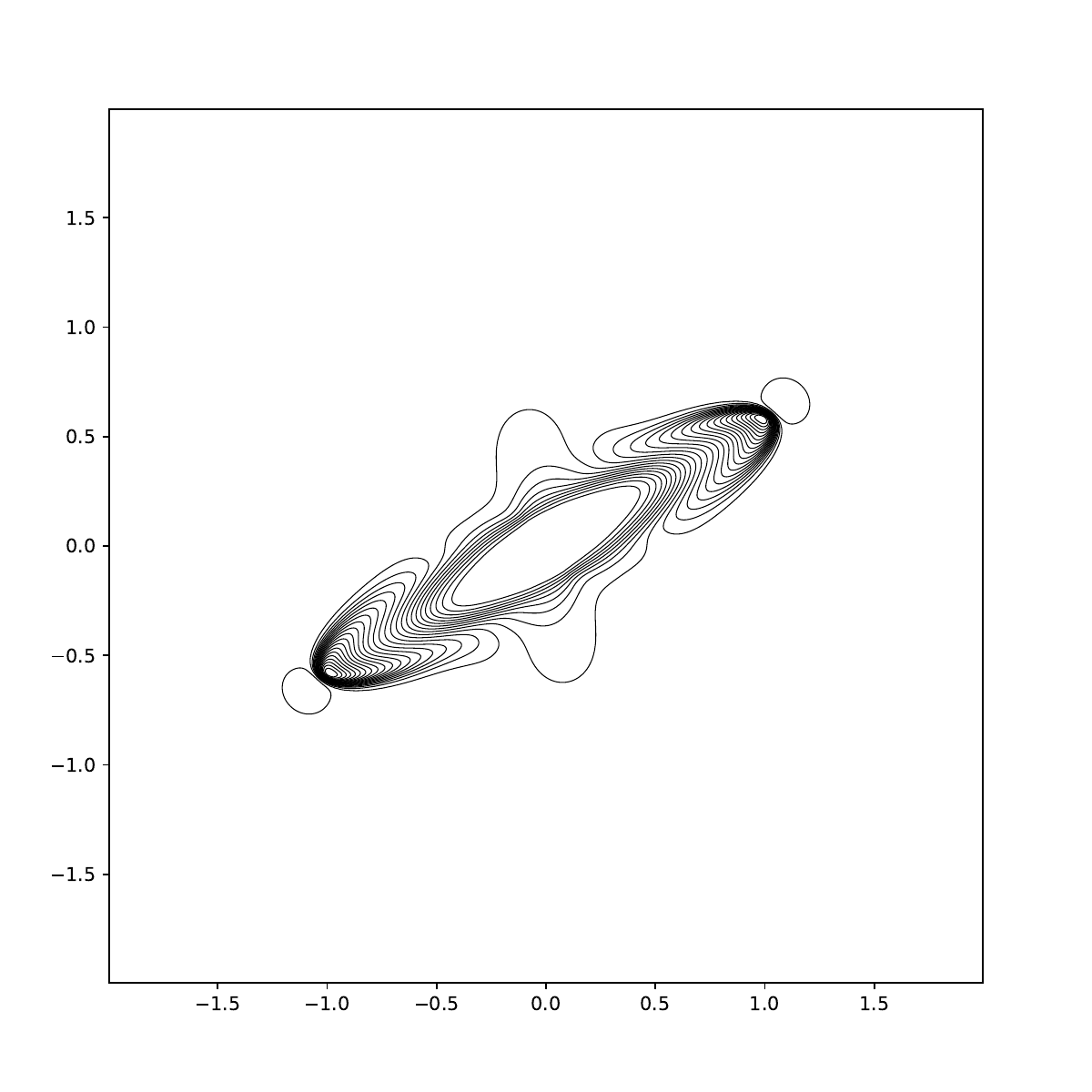}
 }
 \subfigure[\label{fig:t02_density_HLLX}{Density (HLLx)}]{
 \includegraphics[width=0.48\textwidth]{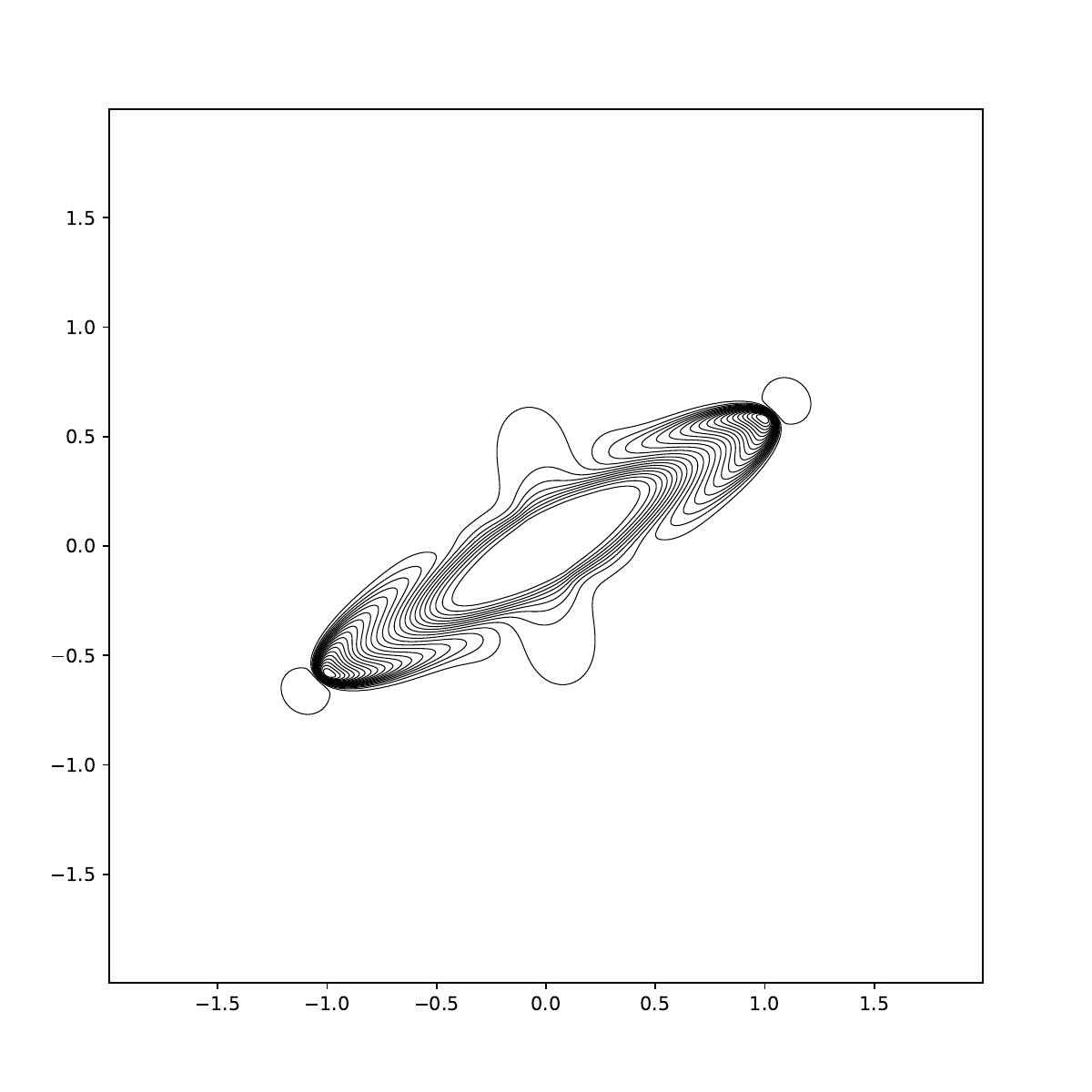}
 } 
\subfigure[\label{fig:t02_total_pressure_HLLD}{Total pressure (HLLD)}]{
 \includegraphics[width=0.48\textwidth]{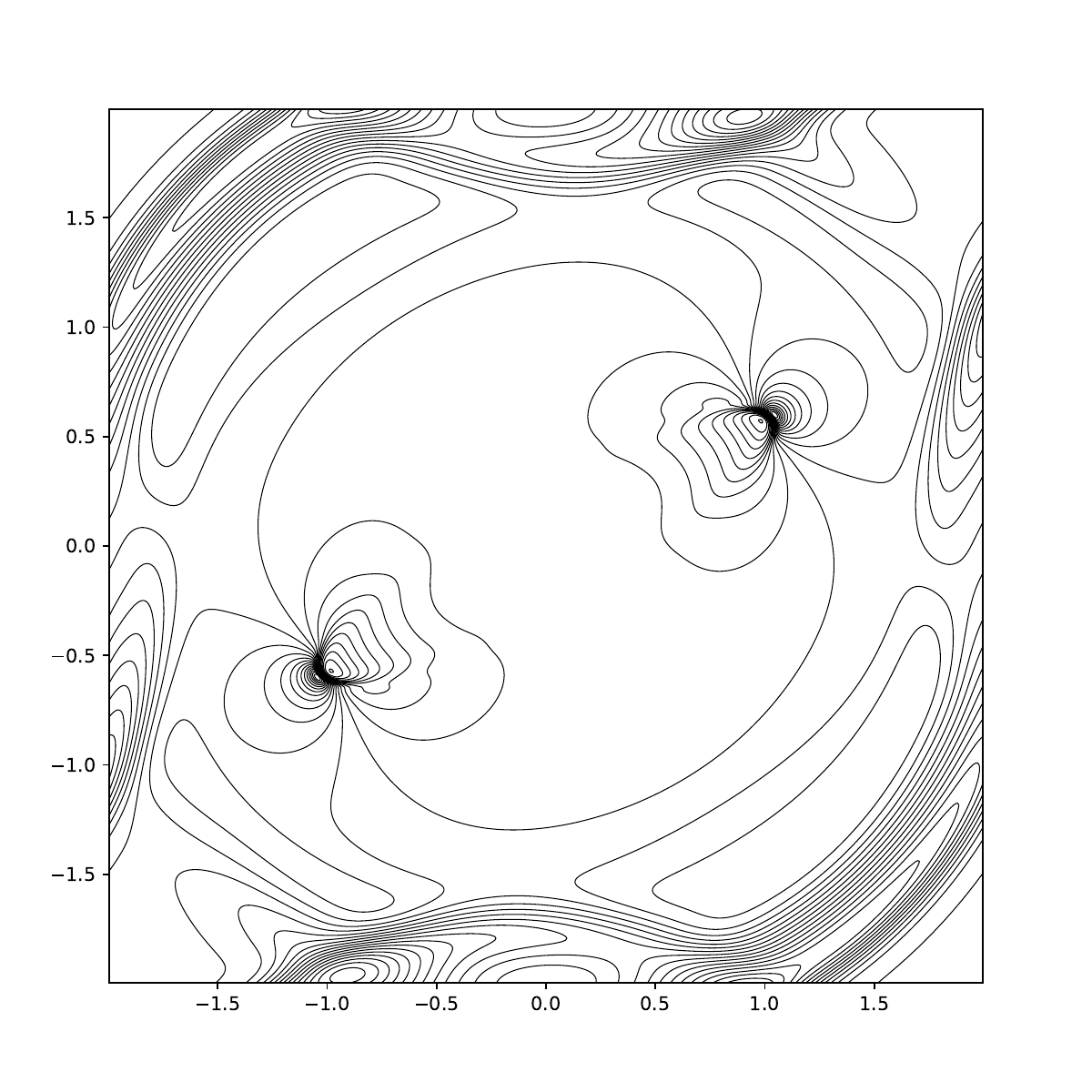}
 }
 \subfigure[\label{fig:t02_total_pressure_HLLX}{Total pressure (HLLx)}]{
 \includegraphics[width=0.48\textwidth]{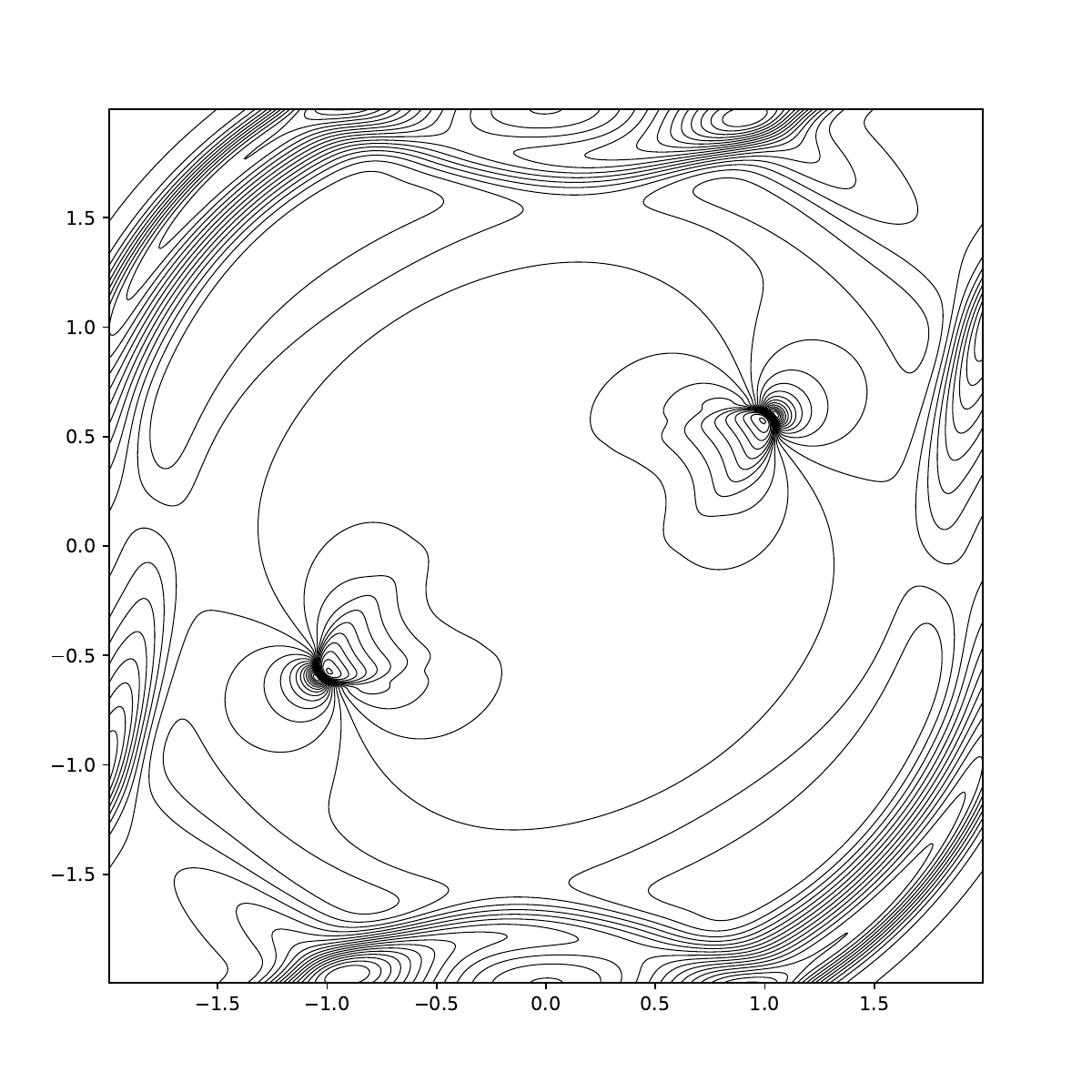}
 } 
 \caption{Results of the MHD blast problem.  In each plot, 30 contour lines are shown for density $\rho\in[0.16, 2.83]$ and for total pressure $P\in[46.5, 55.9]$, approximately.}
 \label{fig:blast_t02}
\end{figure}

\begin{figure}[h]
 \centering
 \subfigure[\label{fig:blast_density_slices_vertical}{}]{
 \includegraphics[width=0.48\textwidth]{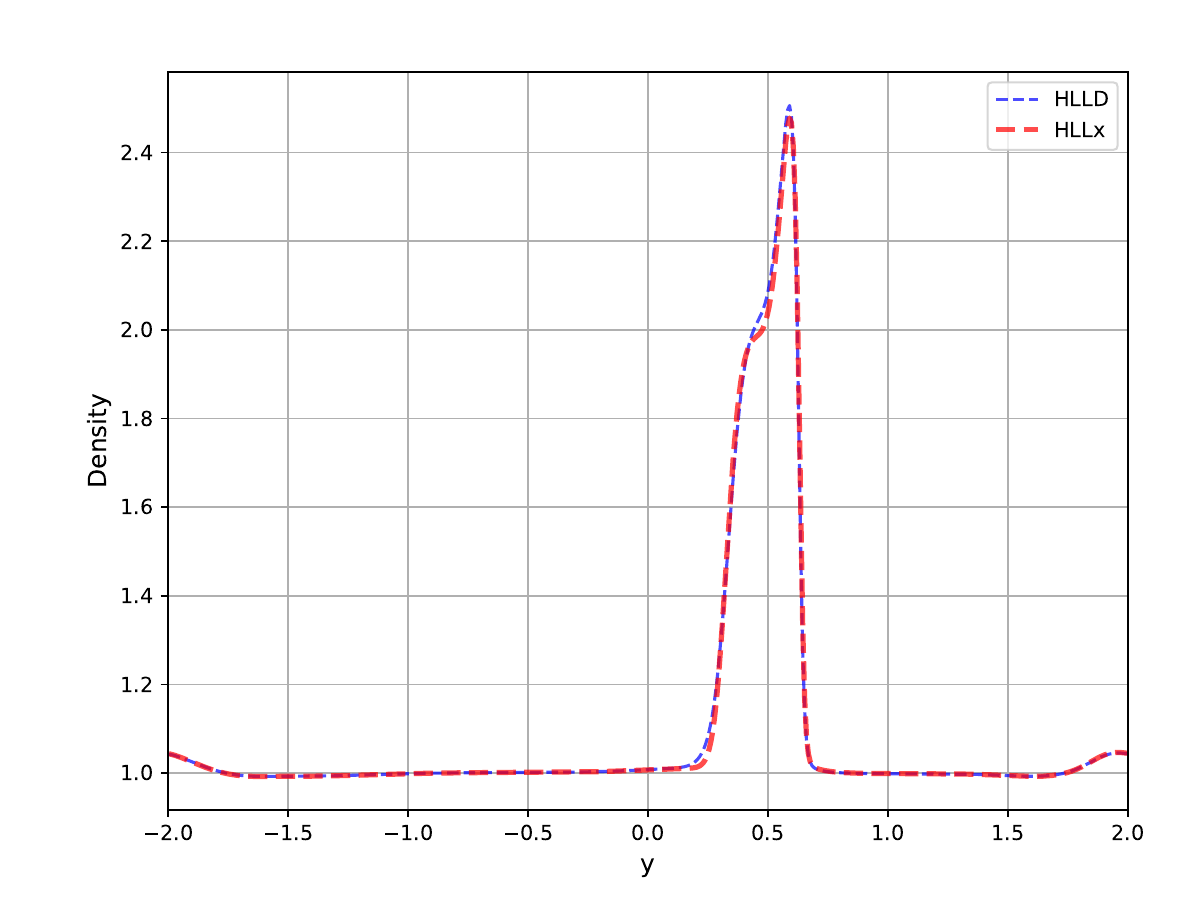}
 }
 \subfigure[\label{fig:blast_total_pressure_slices_vertical}{}]{
 \includegraphics[width=0.48\textwidth]{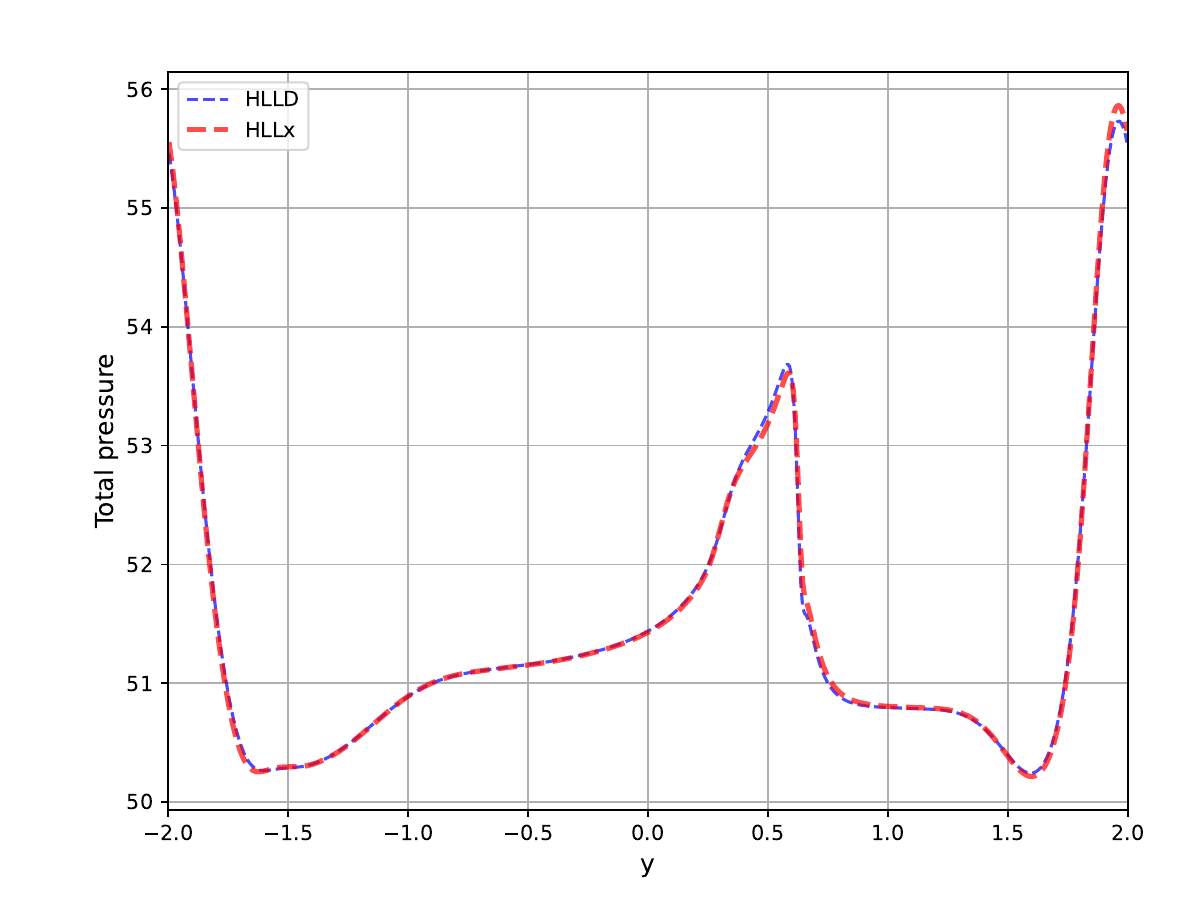}
 }  
 \caption{A vertical slice (at $x=0.9$) of the MHD blast problem.}
 \label{fig:blast_slices}
\end{figure}

In this case, we simulate the propagation of MHD shocks in a
magnetised medium, which is a stringent problem and widely used test case \cite{BALSARA1999,Londrillo_2000}. In this problem, a high-pressure area exists in the middle of the computational domain, and the high pressure drives wave structures under a strong magnetic field. Specifically, here we use the same initial conditions as in Ref. \cite{Minoshima2020}. The ambient conditions are 
\begin{equation}
(\rho, u, v, w, p, B_x, B_y, B_z)=(1, 0, 0, 0, 1, B_0\cos(\theta), B_0\sin(\theta), 0),
\end{equation}
\noindent where $B_0=10$ and $\theta=30^{\circ}$, on a $[-2,2]\times[-2,2]$ domain. Additionally, the high-pressure area is within $\sqrt{x^2+y^2}<0.125$, where the thermal pressure is 100. Here we use the same discretisation and numerical schemes as in the last test case. Numerical results at $t=0.2$ are shown in Figures \ref{fig:blast_t02} and \ref{fig:blast_slices}. 

In Figure \ref{fig:blast_t02}, we can see that strong compressible waves propagate parallel to the magnetic field. In the ambient plasma, the Alfv\'en speed along the magnetic field is faster than the sound speed, but the opposite is true within the high thermal pressure cylinder. Therefore, the shocks here cannot be simply categorised as slow shocks.
We observe that the present HLLx scheme captures sharper structures, more obviously in the density distribution. Nonetheless, the differences in this test case should not be overstated, as the condition used limits the importance of the slow mode. 

A stronger magnetic field may lead to larger differences, but it would also challenge the robustness of both HLLD and HLLx schemes \cite{TREMBLIN2024,zhang2026}.  Specifically, when using $B_0=100$, negative thermal pressure soon develops. The overall result of the HLLD scheme deteriorates, exhibiting spurious oscillations. The HLLx scheme also relies on the HLLD-ec scheme in regions where negative thermal pressure exists, which, as has been mentioned, is however inevitable when the slow speed cannot be  properly calculated. Note that in previous test cases the HLLD-ec scheme is not activated.

\section{Concluding remarks}
 
 To further exploit the potential of the HLL-type approximate Riemann solution for the MHD equations, we have included the slow magnetoacoustic mode when designing the numerical flux.  In doing so, we have specifically addressed the following issues:

 (I) We have shown that if the longitudinal velocity and the total pressure between the slow waves within the Riemann are provided, we may in fact solve a well-posed problem to obtain the other intermediate states in the whole Riemann fan, assuming a fully developed seven-wave configuration for the approximate Riemann solution.

(II) We have systematically discussed the degeneracy of the MHD Riemann solution.  We show that the five-wave configuration used by the classic HLLD scheme is one of the degenerate scenarios. For the possible degenerate scenarios, we have provided the conditions for the problems to be well-posed.

 (III) Inspired by the scenario in which the slow mode is the only compressible mode,  we have designed an "equivalent" wave speed taking into account the varying compressibility of the fast and slow magnetoacoustic modes, for estimating the intermediate longitudinal velocity and total pressure between the left- and right-going slow waves. The resulting estimates make the seven-wave configuration solvable and, more importantly, effectively connect the fully developed seven-wave configuration to the degenerate scenarios.

(IV) The new multi-state HLL-type scheme, denoted as HLLx, is tested in various classic numerical cases. In particular, the new scheme is as accurate as the Roe scheme when the slow mode is dominant, without risking entropy-condition violations when resolving strong expansions. On the other hand, when the slow mode is negligible, the present scheme achieves accuracy slightly better than the classic HLLD scheme. 

Finally, we should emphasise the importance of improving robustness, or more specifically, the ability to preserve the positivity of scalar variables. This is especially important for a method that needs to estimate the slow magnetoacoustic speed, which can hardly tolerate negative pressure in numerical solutions. Therefore, more effort is needed in this regard.

\section*{Acknowledgments}  
The Research Council of Norway supported FZ through its Centres of Excellence scheme, project number 262622. The results were obtained in the framework of the projects FA9550-18-1-0093 (AFOSR), C16/24/010 (C1 project Internal Funds KU Leuven), G0B5823N and G002523N (FWO Vlaanderen), and 4000145223 (SIDC Data Exploitation, ESA Prodex). SP also acknowledges support from the Open SESAME project, which has received funding from the Horizon Europe programme (ERC-AdG agreement No.\ 101141362).

\section*{Declaration of generative AI and AI-assisted technologies in the manuscript preparation process} 

During the preparation of this work, the authors used Claude Sonnet 5, Anthropic, to assist with editing Python visualization scripts used to generate the plots of the numerical results discussed in this manuscript.  After using the AI tool, the authors reviewed and edited the content as needed and take full responsibility for the content of the published article.

\bibliography{ref_MF}

@article{Porth_2014,
doi = {10.1088/0067-0049/214/1/4}, 
year = {2014},
month = {aug},
publisher = {The American Astronomical Society},
volume = {214},
number = {1},
pages = {4},
author = {Porth, O. and Xia, C. and Hendrix, T. and Moschou, S. P. and Keppens, R.},
title = {MPI-AMRVAC FOR SOLAR AND ASTROPHYSICS},
journal = {The Astrophysical Journal Supplement Series},
}

@article{Snow2021OT,
  author  = {Snow, Ben and Hillier, Andrew and Murtas, Giulia and Botha, Gert J. J.},
  title   = {Shock identification and classification in {2D} magnetohydrodynamic compressible turbulence—Orszag–Tang vortex},
  journal = {Experimental Results},
  year    = {2021},
  volume  = {2},
  pages   = {e35},
  doi     = {10.1017/exp.2021.28},
  publisher = {Cambridge University Press}
}

@article{Londrillo_2000,
doi = {10.1086/308344}, 
year = {2000},
month = {feb},
publisher = {},
volume = {530},
number = {1},
pages = {508},
author = {Londrillo, P. and Del Zanna, L.},
title = {High-Order Upwind Schemes for
Multidimensional Magnetohydrodynamics},
journal = {The Astrophysical Journal}, 
}

@article{HARTEN1983_2,
title = {Self adjusting grid methods for one-dimensional hyperbolic conservation laws},
journal = {Journal of Computational Physics},
volume = {50},
number = {2},
pages = {235-269},
year = {1983},
issn = {0021-9991},
doi = {10.1016/0021-9991(83)90066-9}, 
author = {Ami Harten and James M Hyman}, 
}

@Article{Balsara1998,
  author    = {Dinshaw S. Balsara},
  journal   = {The Astrophysical Journal Supplement Series},
  title     = {Linearized Formulation of the Riemann Problem for Adiabatic and Isothermal Magnetohydrodynamics},
  year      = {1998},
  month     = {may},
  number    = {1},
  pages     = {119--131},
  volume    = {116},
  doi       = {10.1086/313092},
  publisher = {American Astronomical Society},
}

@article{BALSARA1999,
title = {A Staggered Mesh Algorithm Using High Order {G}odunov Fluxes to Ensure Solenoidal Magnetic Fields in Magnetohydrodynamic Simulations},
journal = {Journal of Computational Physics},
volume = {149},
number = {2},
pages = {270-292},
year = {1999},
issn = {0021-9991},
doi = {10.1006/jcph.1998.6153},
author = {Dinshaw S Balsara and Daniel S Spicer},
}

@article{TREMBLIN2024,
Author ={Pascal Tremblin and  R\'emi Bourgeois and 
Sol\`ene Bulteau and
Samuel Kokh and Thomas Padioleau and 
Maxime Delorme and Antoine Strugarek and
 Matthias Gonz\'alez and Allan Sacha Brun},
title = {A multi-dimensional, robust, and cell-centered finite-volume scheme for the ideal {MHD} equations},
journal = {Journal of Computational Physics},
volume = {519},
pages = {113455},
year = {2024},
issn = {0021-9991},
doi = {10.1016/j.jcp.2024.113455}, 
}

@article{Wu_2018,
author = {Wu, Kailiang},
title = {Positivity-Preserving Analysis of Numerical Schemes for Ideal Magnetohydrodynamics},
journal = {SIAM Journal on Numerical Analysis},
volume = {56},
number = {4},
pages = {2124-2147},
year = {2018},
doi = {10.1137/18M1168017}, 
}

@Book{Goedbloed2004,
  Title                    = {Principles of Magnetohydrodynamics: With Applications to Laboratory and Astrophysical Plasmas},
  Author                   = {Goedbloed, J. P. Hans and Poedts, Stefaan},
  Publisher                = {Cambridge University Press},
  Year                     = {2004},

  Doi                      = {10.1017/CBO9780511616945},
  Place                    = {Cambridge}
}

@Article{Gurski2004,
  author    = {Gurski, K. F.},
  journal   = {SIAM Journal on Scientific Computing},
  title     = {An {HLLC}-Type Approximate {R}iemann Solver for Ideal Magnetohydrodynamics},
  year      = {2004},
  number    = {6},
  pages     = {2165-2187},
  volume    = {25},
  doi       = {10.1137/S1064827502407962}, 
}

@article{Zhang2024,
author = {Zhang, Fan},
title = {Improving the quantification of overshooting shock-capturing oscillations},
journal = {Progress in Computational Fluid Dynamics},
doi = {10.1504/PCFD.2024.138236}, 
volume = {24},
number = {3},
pages = {135-142},
year = {2024},
}

@article{Minoshima_2019,
doi = {10.3847/1538-4365/ab1a36},
year = {2019},
month = {may},
publisher = {The American Astronomical Society},
volume = {242},
number = {2},
pages = {14},
author = {Takashi Minoshima and Takahiro Miyoshi and Yosuke Matsumoto},
title = {A High-order Weighted Finite Difference Scheme with a Multistate Approximate {R}iemann Solver for Divergence-free Magnetohydrodynamic Simulations},
journal = {The Astrophysical Journal Supplement Series},
}

@Article{Perri2022,
  author    = {Barbara Perri and Peter Leitner and Michaela Brchnelova and Tinatin Baratashvili and B{\l}a{\.{z}}ej Ku{\'{z}}ma and Fan Zhang and Andrea Lani and Stefaan Poedts},
  journal   = {The Astrophysical Journal},
  title     = {{COCONUT}, a novel fast-converging {MHD} model for solar corona simulations: {I}. Benchmarking and optimization of polytropic solutions},
  year      = {2022},
  number    = {1},
  pages     = {19},
  volume    = {936},
  doi       = {10.3847/1538-4357/ac7237},
}

@Article{Li2005,
  Title                    = {An {HLLC} {R}iemann solver for magneto-hydrodynamics},
  Author                   = {Shengtai Li},
  Journal                  = {Journal of Computational Physics},
  Year                     = {2005},
  Number                   = {1},
  Pages                    = {344-357},
  Volume                   = {203},

  Doi                      = {10.1016/j.jcp.2004.08.020},
  ISSN                     = {0021-9991}
}

@Article{Minoshima2020,
  Title                    = {A Multistate Low-dissipation Advection Upstream Splitting Method for Ideal Magnetohydrodynamics},
  Author                   = {Takashi Minoshima and Keiichi Kitamura and Takahiro Miyoshi},
  Journal                  = {The Astrophysical Journal Supplement Series},
  Year                     = {2020},

  Month                    = {may},
  Number                   = {1},
  Pages                    = {12},
  Volume                   = {248},

  Doi                      = {10.3847/1538-4365/ab8aee},
  Publisher                = {American Astronomical Society}
}

@Article{Miyoshi2005,
  Title                    = {A multi-state {HLL} approximate {R}iemann solver for ideal magnetohydrodynamics},
  Author                   = {Takahiro Miyoshi and Kanya Kusano},
  Journal                  = {Journal of Computational Physics},
  Year                     = {2005},
  Number                   = {1},
  Pages                    = {315-344},
  Volume                   = {208},

  Doi                      = {10.1016/j.jcp.2005.02.017},
  ISSN                     = {0021-9991}
}

@Article{Popovas2025,
  Title                    = {{DISPATCH} methods: an approximate, entropy-based {R}iemann solver for ideal magnetohydrodynamics},
  Author                   = {Andrius Popovas},
  Journal                  = {Astronomy \& Astrophysics},
  Year                     = {2025},    
  Volume                   = {698},
  Pages                    = {A69},
  Doi                      = {10.1051/0004-6361/202554028},
}

@Article{Toro1994,
  Title                    = {Restoration of the contact surface in the {HLL}-{R}iemann solver},
  Author                   = {E. F. Toro and M. Spruce and W. Speares},
  Journal                  = {Shock Waves},
  Year                     = {1994},
  Pages                    = {25-34},
  Volume                   = {4},

  Doi                      = {10.1007/BF01414629}
}

@article{Roe1981,
title = {Approximate {R}iemann solvers, parameter vectors, and difference schemes},
journal = {Journal of Computational Physics},
volume = {43},
number = {2},
pages = {357-372},
year = {1981},
issn = {0021-9991},
doi = {10.1016/0021-9991(81)90128-5}, 
author = {P.L Roe}, 
}

@article{Roe_1996,
author = {Roe, P. L. and Balsara, D. S.},
title = {Notes on the Eigensystem of Magnetohydrodynamics},
journal = {SIAM Journal on Applied Mathematics},
volume = {56},
number = {1},
pages = {57-67},
year = {1996},
doi = {10.1137/S003613999427084X}, 
}

@article{BRIO1988,
title = {An upwind differencing scheme for the equations of ideal magnetohydrodynamics},
journal = {Journal of Computational Physics},
volume = {75},
number = {2},
pages = {400-422},
year = {1988},
issn = {0021-9991},
doi = {10.1016/0021-9991(88)90120-9}, 
author = {M Brio and C.C Wu}, 
}

@article{DAI1994,
title = {An Approximate {R}iemann Solver for Ideal Magnetohydrodynamics},
journal = {Journal of Computational Physics},
volume = {111},
number = {2},
pages = {354-372},
year = {1994},
issn = {0021-9991},
doi = {10.1006/jcph.1994.1069}, 
author = {Wenlong Dai and Paul R. Woodward}, 
}

@Article{Dedner_2002,
  author    = {A. Dedner and F. Kemm and D. Kröner and C.-D. Munz and T. Schnitzer and M. Wesenberg},
  journal   = {Journal of Computational Physics},
  title     = {Hyperbolic Divergence Cleaning for the {MHD} Equations},
  year      = {2002},
  month     = {jan},
  number    = {2},
  pages     = {645--673},
  volume    = {175},
  doi       = {10.1006/jcph.2001.6961},
  publisher = {Elsevier {BV}},
}

@Article{Mignone_2007,
  author    = {A. Mignone and G. Bodo and S. Massaglia and T. Matsakos and O. Tesileanu and C. Zanni and A. Ferrari},
  journal   = {The Astrophysical Journal Supplement Series},
  title     = {{PLUTO}: A Numerical Code for Computational Astrophysics},
  year      = {2007},
  month     = {may},
  number    = {1},
  pages     = {228--242},
  volume    = {170},
  doi       = {10.1086/513316},
  publisher = {American Astronomical Society},
}

@article{Navarro_2025,
	author = {Navarro, A. and Khomenko, E. and Vitas, N. and Felipe, T.},
	title = {Modeling solar atmosphere dynamics with MAGEC},
	DOI= "10.1051/0004-6361/202556221",
	journal = {Astronomy \& Astrophysics},
	year = 2025,
	volume = 703,
	pages = "A298",
}

@article{Isaacson_1992,
author = {Isaacson, Eli and Temple, Blake},
title = {Nonlinear Resonance in Systems of Conservation Laws},
journal = {SIAM Journal on Applied Mathematics},
volume = {52},
number = {5},
pages = {1260-1278},
year = {1992},
doi = {10.1137/0152073}, 
}

@article{Orszag1979, 
title={Small-scale structure of two-dimensional magnetohydrodynamic turbulence}, volume={90}, 
DOI={10.1017/S002211207900210X}, 
number={1}, 
journal={Journal of Fluid Mechanics}, 
author={Orszag, Steven A. and Tang, Cha-Mei}, 
year={1979}, 
pages={129–143}
}

@Article{Toth_2000,
  author    = {G{\'{a}}bor T{\'{o}}th},
  journal   = {Journal of Computational Physics},
  title     = {The $\nabla$$\cdotp${B}=0 Constraint in Shock-Capturing Magnetohydrodynamics Codes},
  year      = {2000},
  month     = {jul},
  number    = {2},
  pages     = {605--652},
  volume    = {161},
  doi       = {10.1006/jcph.2000.6519},
  publisher = {Elsevier {BV}},
}

@article{EINFELDT1991,
title = {On {G}odunov-type methods near low densities},
journal = {Journal of Computational Physics},
volume = {92},
number = {2},
pages = {273-295},
year = {1991},
issn = {0021-9991},
doi = {10.1016/0021-9991(91)90211-3}, 
author = {B Einfeldt and C.D Munz and P.L Roe and B Sjögreen}, 
}

@article{DUMBSER2016,
title = {A new efficient formulation of the {HLLEM} {R}iemann solver for general conservative and non-conservative hyperbolic systems},
journal = {Journal of Computational Physics},
volume = {304},
pages = {275-319},
year = {2016},
issn = {0021-9991},
doi = {10.1016/j.jcp.2015.10.014}, 
author = {Michael Dumbser and Dinshaw S. Balsara}, 
}

@Article{Harten_1983,
  author    = {Amiram Harten and Peter D. Lax and Bram van Leer},
  journal   = {{SIAM} Review},
  title     = {On Upstream Differencing and {G}odunov-Type Schemes for Hyperbolic Conservation Laws},
  year      = {1983},
  month     = {jan},
  number    = {1},
  pages     = {35--61},
  volume    = {25},
  doi       = {10.1137/1025002},
  publisher = {Society for Industrial {\&} Applied Mathematics ({SIAM})},
}

@Article{Einfeldt_1988,
  author    = {Bernd Einfeldt},
  journal   = {{SIAM} Journal on Numerical Analysis},
  title     = {On {G}odunov-Type Methods for Gas Dynamics},
  year      = {1988},
  month     = {apr},
  number    = {2},
  pages     = {294--318},
  volume    = {25},
  doi       = {10.1137/0725021},
  publisher = {Society for Industrial {\&} Applied Mathematics ({SIAM})},
}

@Article{Batten_1997,
  author    = {P. Batten and N. Clarke and C. Lambert and D. M. Causon},
  journal   = {{SIAM} Journal on Scientific Computing},
  title     = {On the Choice of Wavespeeds for the {HLLC} {R}iemann Solver},
  year      = {1997},
  month     = {nov},
  number    = {6},
  pages     = {1553--1570},
  volume    = {18},
  doi       = {10.1137/s1064827593260140},
  publisher = {Society for Industrial {\&} Applied Mathematics ({SIAM})},
}

@Book{Toro2009,
  author    = {Eleuterio F. Toro},
  publisher = {Springer},
  title     = {Riemann Solvers and Numerical Methods for Fluid Dynamics},
  year      = {2009},
  edition   = {Third},
}

@article{Mattia2021,
    author = {Mattia, G and Mignone, A},
    title = "{A comparison of approximate non-linear Riemann solvers for Relativistic MHD}",
    journal = {Monthly Notices of the Royal Astronomical Society},
    volume = {510},
    number = {1},
    pages = {481-499},
    year = {2022},
    month = {11}, 
    issn = {0035-8711},
    doi = {10.1093/mnras/stab3373},  
}

@article{Balsara_2004,
doi = {10.1086/381051}, 
year = {2004},
month = {feb},
publisher = {},
volume = {602},
number = {2},
pages = {1079},
author = {Balsara, Dinshaw S. and Kim, Jongsoo},
title = {A Comparison between Divergence-Cleaning and Staggered-Mesh Formulations for Numerical Magnetohydrodynamics},
journal = {The Astrophysical Journal},
}

@book{Priest2014,
  author    = {Priest, Eric},
  title     = {Magnetohydrodynamics of the {Sun}},
  publisher = {Cambridge University Press},
  place   = {Cambridge},
  year      = {2014},
  isbn      = {978-0-521-85471-9},
  doi       = {10.1017/CBO9781139020732}
}

@ARTICLE{zhang2026,
title = {On energy consistency of intermediate states in {HLL}-type {MHD} {R}iemann solvers},
journal = {Journal of Computational Physics},
volume = {553},
pages = {114724},
year = {2026},
issn = {0021-9991},
doi = {10.1016/j.jcp.2026.114724}, 
author = {Fan Zhang and Andrea Lani and Stefaan Poedts},
}

@article{Mignone_2006,
    author = {Mignone, A. and Bodo, G.},
    title = {An HLLC Riemann solver for relativistic flows – II. Magnetohydrodynamics},
    journal = {Monthly Notices of the Royal Astronomical Society},
    volume = {368},
    number = {3},
    pages = {1040-1054},
    year = {2006},
    month = {04}, 
    issn = {0035-8711},
    doi = {10.1111/j.1365-2966.2006.10162.x}, 
}

@article{Mignone_2009,
    author = {Mignone, A. and Ugliano, M. and Bodo, G.},
    title = {A five-wave Harten–Lax–van Leer Riemann solver for relativistic magnetohydrodynamics},
    journal = {Monthly Notices of the Royal Astronomical Society},
    volume = {393},
    number = {4},
    pages = {1141-1156},
    year = {2009},
    month = {03}, 
    issn = {0035-8711},
    doi = {10.1111/j.1365-2966.2008.14221.x}, }

@ARTICLE{White_2016,
       author = {{White}, Christopher J. and {Stone}, James M. and {Gammie}, Charles F.},
        title = "{An Extension of the Athena++ Code Framework for GRMHD Based on Advanced Riemann Solvers and Staggered-mesh Constrained Transport}",
      journal = {The Astrophysical Journal Supplement Series}, 
         year = 2016,
        month = aug,
       volume = {225},
       number = {2},
          eid = {22},
        pages = {22},
          doi = {10.3847/0067-0049/225/2/22},  
}

@ARTICLE{Evans_1988,
       author = {{Evans}, Charles R. and {Hawley}, John F.},
        title = "{Simulation of Magnetohydrodynamic Flows: A Constrained Transport Model}",
      journal = {The Astrophysical Journal}, 
         year = 1988,
        month = sep,
       volume = {332},
        pages = {659},
          doi = {10.1086/166684}, 
}
\end{document}